\documentclass[twocolumn,twocolappendix]{aastex702}
\usepackage{mathrsfs}
\usepackage{subfigure}
\usepackage{epstopdf}
\usepackage{amsmath}
\usepackage{longtable}
\usepackage{booktabs}
\usepackage{hyperref} \usepackage{threeparttable}
\usepackage{graphicx} 
\usepackage{overpic}
\def\etal {et al.~}

\newbox\grsign \setbox\grsign=\hbox{$>$} \newdimen\grdimen \grdimen=\ht\grsign
\newbox\laxbox \newbox\gaxbox
\setbox\gaxbox=\hbox{\raise.5ex\hbox{$>$}\llap
     {\lower.5ex\hbox{$\sim$}}}\ht1=\grdimen\dp1=0pt
\setbox\laxbox=\hbox{\raise.5ex\hbox{$<$}\llap
     {\lower.5ex\hbox{$\sim$}}}\ht2=\grdimen\dp2=0pt

\shortauthors{Lin \etal}
\graphicspath{{figs/}}

\definecolor{malachite}{rgb}{0.34, 0.7, 0.22}

\begin{document}

\title{Large-Scale Spiral Structure of the Milky Way Traced by Young Classical Cepheids}


\author{Z. H. Lin}
\affiliation{Purple Mountain Observatory, Chinese Academy of Sciences, Nanjing 210023, People's Republic of China}
\affiliation{State Key Laboratory of Radio Astronomy and Technology, Purple Mountain Observatory, Chinese Academy of Sciences, 10 Yuanhua Road, Nanjing 210023, China}
\email[show]{linzh@pmo.ac.cn}

\author{C. J. Hao}
\affiliation{Purple Mountain Observatory, Chinese Academy of Sciences, Nanjing 210023, People's Republic of China}
\affiliation{State Key Laboratory of Radio Astronomy and Technology, Purple Mountain Observatory, Chinese Academy of Sciences, 10 Yuanhua Road, Nanjing 210023, China}
\email{cjhao@pmo.ac.cn}

\author{D. J. Liu}
\affiliation{College of Mathematics and Physics, China Three Gorges University, Yichang 443000, People's Republic of China}
\affiliation{Center for Astronomy and Space Sciences, China Three Gorges University, Yichang 443000, People's Republic of China}
\email{liudejian@ctgu.edu.cn}

\author{Y. J. Li}
\affiliation{Purple Mountain Observatory, Chinese Academy of Sciences, Nanjing 210023, People's Republic of China}
\affiliation{State Key Laboratory of Radio Astronomy and Technology, Purple Mountain Observatory, Chinese Academy of Sciences, 10 Yuanhua Road, Nanjing 210023, China}
\affiliation{University of Science and Technology of China, 96 Jinzhai Road, Hefei 230026, People's Republic of China}
\email{liyj@pmo.ac.cn}

\author{J. J. Li}
\affiliation{Purple Mountain Observatory, Chinese Academy of Sciences, Nanjing 210023, People's Republic of China}
\affiliation{State Key Laboratory of Radio Astronomy and Technology, Purple Mountain Observatory, Chinese Academy of Sciences, 10 Yuanhua Road, Nanjing 210023, China}
\affiliation{University of Science and Technology of China, 96 Jinzhai Road, Hefei 230026, People's Republic of China}
\email{jjli@pmo.ac.cn}

\author{Y. W. Dong}
\affiliation{Purple Mountain Observatory, Chinese Academy of Sciences, Nanjing 210023, People's Republic of China}
\affiliation{State Key Laboratory of Radio Astronomy and Technology, Purple Mountain Observatory, Chinese Academy of Sciences, 10 Yuanhua Road, Nanjing 210023, China}
\affiliation{University of Science and Technology of China, 96 Jinzhai Road, Hefei 230026, People's Republic of China}
\email{dongyw@pmo.ac.cn}

\author{Y. X. He}
\affiliation{State Key Laboratory of Radio Astronomy and Technology, Xinjiang Astronomical Observatory, CAS, 150 Science 1-Street, Urumqi, Xinjiang 830011, 
P. R. China }
\affiliation{ University of the Chinese Academy of Sciences, Beijing 100080, P. R. China}
\affiliation{ Xinjiang Key Laboratory of Radio Astrophysics, Urumqi 830011, P. R. China}
\email{heyuxin@xao.ac.cn}

\author{Jarken Esimbek}
\affiliation{State Key Laboratory of Radio Astronomy and Technology, Xinjiang Astronomical Observatory, CAS, 150 Science 1-Street, Urumqi, Xinjiang 830011, 
P. R. China }
\affiliation{ University of the Chinese Academy of Sciences, Beijing 100080, P. R. China}
\affiliation{ Xinjiang Key Laboratory of Radio Astrophysics, Urumqi 830011, P. R. China}
\email{jarken@xao.ac.cn}

\begin{abstract}
Using 985 classical Cepheids younger than 100~Myr, we investigate the large-scale spiral structure of the Milky Way by comparing their spatial and kinematic distributions with the R19 (four-arm) and X23 (multiple-arm) spiral-arm models. Both models provide reasonably good descriptions of the current sample: R19 yields slightly smaller point-to-arm distances, whereas X23 has a higher supported-arm fraction, but the descriptive statistics do not provide a decisive preference between them. Model-dependent arm assignment and refitting suggest that the distribution of the youngest Classical Cepheids (20--60~Myr)
is consistent with an X23-like multiple-arm in which the Perseus and Norma arms form two dominant inner arms extending toward the ends of the Galactic bar, the Sagittarius and Centaurus segments are possible branches of these inner arms, and Local, Outer, and other features constitute additional arm segments. The updated model improves the constraints on the southern spiral structure, yielding smaller point-to-arm distances and a higher supported-arm fraction. Some of these structures extend to heliocentric distances of approximately 15~kpc.
\end{abstract}

\keywords{\uat{Galaxy structure}{622} --- \uat{Milky Way Galaxy}{1054)} --- \uat{Cepheid variable stars}{218}}

\section{Introduction}

Since the first identification of spiral structure in the Milky Way by \citet{morgan1953}, accurately mapping its morphology has remained a long-standing challenge. Numerous theoretical models have been proposed \citep{steiman2010}, yet the true structure of the Galaxy remains uncertain, largely because precise distances to spiral-arm tracers were historically lacking. As spiral arms are the main sites for star formation, accurate distances to objects in star-forming regions and young stellar tracers provide important constraints on the Galactic spiral structure \citep{xu2018b,xu2026}.

By using the parallax measurement method, high-precision distance determination of spiral-arm tracers were obtained, marking the beginning of the era of precise mapping of the Milky Way spiral structure~\citep{xu2006}. Subsequent large astrometric programs, including the Bar and Spiral Structure Legacy (BeSSeL) survey \citep{brunthaler2011} and the VLBI Exploration of Radio Astrometry (VERA) array \citep{vera2020}, have systematically advanced this effort. Based on high-precision parallax measurements of nearly 200 masers associated with high-mass star-forming regions, \citet{reid2019} delineated a four-arm spiral model of the Milky Way, hereafter R19 model. 

Using VVV near-infrared distances for a small sample of 50 classical
Cepheids located on the far side of the Galactic disk,
\citet{minniti2021} carried out an exploratory test of whether these
distant tracers were consistent with the maser-defined spiral-arm
framework. Within the context of the R19 model, their results suggested
that the Perseus Arm may extend inward on the far side of the Galaxy
and connect with the Norma Arm. This interpretation provided
preliminary evidence for a configuration in which two dominant inner
arms branch into four arms beyond Galactocentric radii of approximately
5--6~kpc.

More recently, \citet{xu2023} reexamined the maser distribution and
combined the precise locations of young tracers with spiral-arm tangent
constraints. Departing from the conventional picture of four
continuous arms, they introduced a new multiple-arm framework for the
Milky Way, hereafter the X23 model, in which two dominant inner arms
develop into multiple arm structures in the outer disk. The model proposes that, within the conventional four-arm framework, the Sagittarius and Perseus Arms converge and connect on the far side
of the first Galactic quadrant, while the Norma and Scutum Arms are reinterpreted as parts of a single arm. Consequently, only two dominant spiral arms, the Perseus and Norma Arms, remain in the inner Galaxy. The Perseus and Norma Arms connect directly to the near-side and
far-side ends of the Galactic bar, respectively. This configuration represents a multiple-arm morphology, a common spiral pattern among spiral galaxies~\citep{elmegreen1989,buta2015,wei2024}.

Young massive stars are expected to remain near their birth sites \citep{xu2018a}, and with distance information from \emph{Gaia}, young stellar samples trace spiral structures broadly consistent with those delineated by masers \citep[e.g.,][]{xu2021, hao2021}. The timescale over which stars migrate away from their natal arms remains poorly constrained, with different models predicting distinct behaviors \citep[e.g.,][]{dobbs2014, grand2016}. Limited observational evidence from \citet{ge2024} shows that even B6$-$B7 stars ($\sim$60~Myr, with a broad age spread) trace spiral structures that differ from those defined by O$-$B2 stars by less than the arm width. This observational result suggests that spiral-arm signatures remain imprinted not only in the extremely young massive stars but also in somewhat older populations.

Infrared observations can penetrate the heavily obscured regions of the Galactic disk inaccessible at optical wavelengths, thereby enabling the exploration of larger-scale spiral structures. Classical Cepheids follow a well-defined near-infrared period$-$luminosity relation, enabling reliable distance estimates. 

After correcting for the Galactic warp, \citet{lemasle2022} applied a clustering algorithm mainly to Cepheids younger than $\sim150$ Myr and identified groups broadly consistent with arms traced by masers, OB(A) stars, and Gaia upper-main-sequence overdensities. More recently, \citet{drimmel2025} selected 2\,857 ``dynamically young'' Cepheids and traced density ridges without assigning individual stars to predefined arms. They identified the Perseus and Sagittarius--Carina arms as the most prominent large-scale features and suggested that the Local Arm is more likely a spur, indicating a spiral pattern that may be more open and segmented than conventional grand-design models.

In this Letter, we analyze the spatial and kinematic distributions of young classical Cepheids to test whether they remain in their natal arms or have begun to drift away, and to trace the large-scale spiral structure of the Milky Way.

\section{Sample}
\label{sec2}

\cite{wang2018} established an optimal method to determine extinction ratios and distances of classical Cepheids in the near-infrared, using the mid-infrared intensity-mean magnitude relation and multi-band period$-$luminosity relations. \cite{skowron2025} recalculated distances for 3\,424 classical Cepheids based on mid-infrared photometry from the \textit{Wide-field Infrared Survey Explorer} (WISE) and corrected for (albeit small) extinction effects using a 3D extinction map \citep{bovy2016}. Comparison with a sample of Cepheids having reliable parallax distances demonstrates that the mid-infrared period$-$luminosity distances are accurate to 6\% \citep{skowron2025}.

\begin{figure*}[!ht]
    \centering
    \includegraphics[width=0.8\textwidth]{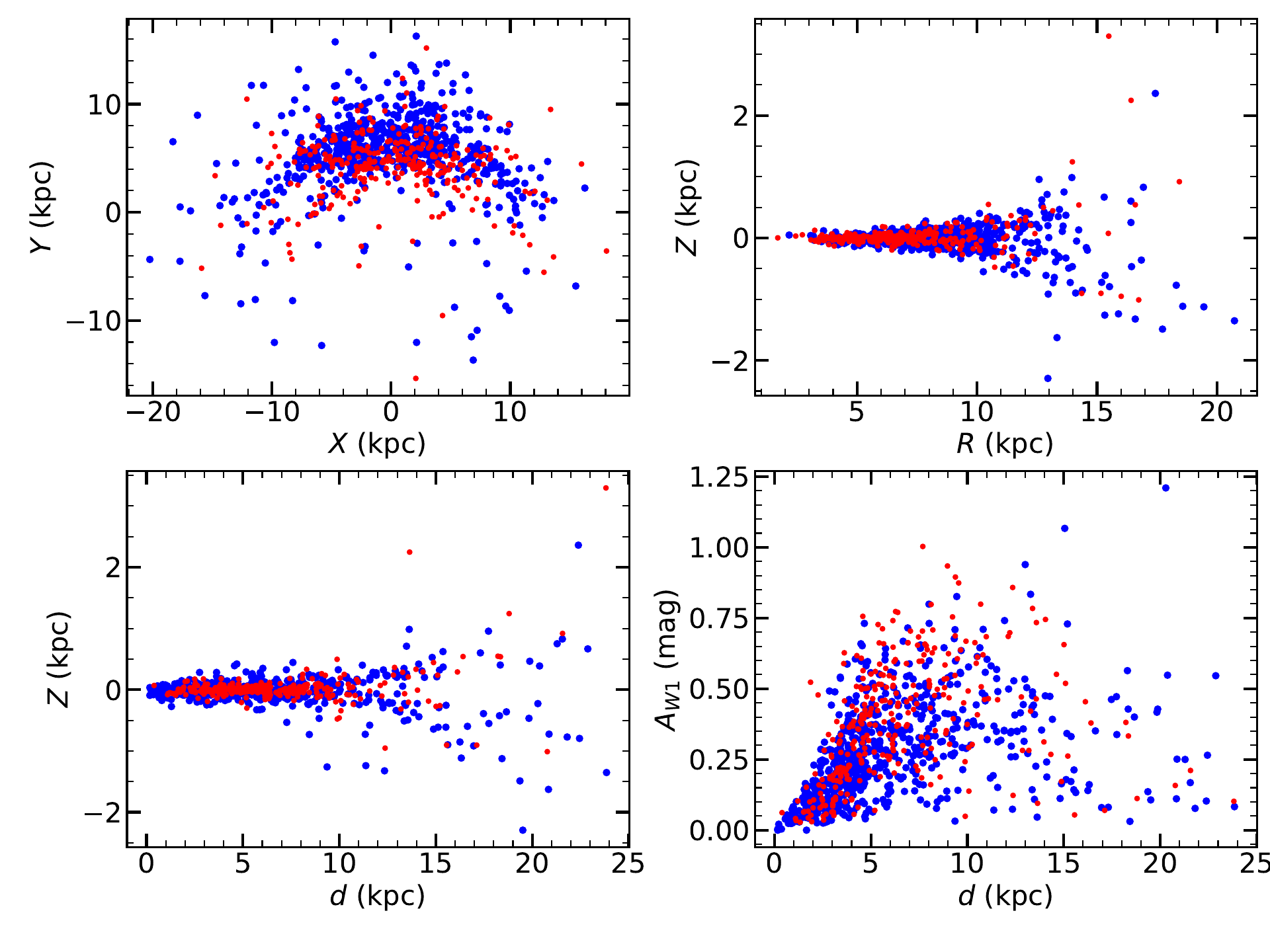}
    \caption{Spatial and extinction distributions of the Classical Cepheid sample. Red and blue points represent subsamples A and B, respectively. The upper-left panel shows the distribution in the Galactic ($X-Y$) plane. The upper-right and lower-left panels show the vertical height ($Z$) as a function of Galactocentric radius ($R$) and heliocentric distance ($d$), respectively. The lower-right panel shows the $W1$-band extinction, ($A_{W1}$), as a function of $d$.}
    \label{figs:densitymap}
\end{figure*}

Classical Cepheids provide reliable age estimates through their period--age relation \citep{bono2005,anderson2016}, although the ages of individual Cepheids may still be affected by factors such as stellar rotation, potentially leading to uncertainties of up to $\sim50\%$. \citet{skowron2025} derived the ages of these Classical Cepheid samples by accounting for both the initial rotation rate and metallicity effects. We assume that the ages are sufficiently reliable to define subsamples of different age ranges and to investigate the spiral structure traced by stellar populations of different ages.

Younger stars are better suited to tracing the process of star formation in spiral arms and their subsequent migration. We therefore select classical Cepheids with ages younger than 100 Myr from the sample of \cite{skowron2025}, yielding a sample of 985 stars. As evolved stars, classical Cepheids typically have ages exceeding $\sim$20~Myr \citep{skowron2025}. To examine potential age-dependent differences in tracing spiral structure while maintaining sufficient sample sizes, we further divide the sample into two age bins: 20--60 Myr (subsample A; 301 stars) and 60--100 Myr (subsample B; 684 stars). 60 Myr is included only in the subsample A to avoid overlap.

The radial velocities of classical Cepheids vary with pulsation phase and may therefore introduce additional uncertainties into measurements obtained at individual epochs~\citep{katz2023,anderson2024}. In this work, we adopt the combined radial velocities provided by the \texttt{radial\_velocity} field in the Gaia DR3 \texttt{gaia\_source} table. These velocities are derived from multiple RVS transits; for sources with $G_{\rm RVS}\leq12$ mag, the published value is the median of the eligible epoch radial velocities, whereas for fainter sources it is determined from the combined cross-correlation functions of the individual transits \citep{katz2023}. Combining multiple transits reduces the sensitivity of the published radial velocity to measurements obtained at any particular pulsation phase. Radial-velocity measurements were available for 655 stars in total, including 149 stars in subsample A and 506 stars in subsample B. Following the formalism presented in the appendix of \citet{reid2009}, we converted the measured heliocentric radial velocities into velocities with respect to the Local Standard of Rest (LSR). In this conversion, we adopted a solar motion of 20 km s$^{-1}$ toward $\alpha_{\rm 1900}=18^{\rm h}$ and $\delta_{\rm 1900}=+30^{\circ}$~\citep{reid2009}.

\section{Result}
\subsection{Spatial Distribution and Consistency with Maser-Traced Spiral-Arm Models}
\label{sec3.1}

\begin{deluxetable*}{lccccccccc}
\tablecaption{Comparison between the R19 and X23 spiral-arm models
\label{tab:model_comparison}}
\tablehead{
\colhead{Statistic} & \multicolumn{3}{c}{All sample} & \multicolumn{3}{c}{Subsample A} & \multicolumn{3}{c}{Subsample B} \\  
\colhead{} & \colhead{R19} & \colhead{X23} & \colhead{New} & \colhead{R19} & \colhead{X23} & \colhead{New} & \colhead{R19} & \colhead{X23} & \colhead{New}
}
\startdata
Median point-to-arm distance, $d_i$ (kpc) & 0.46 & 0.47 & 0.42 & 0.42 & 0.44 & 0.34 & 0.48 & 0.50 & 0.44\\
Median normalized distance, $\delta_i$ & 0.9 & 1.0 & 0.9 & 0.8 & 0.9 & 0.7 & 0.9 & 1.0 & 0.9\\
Mean normalized squared distance, $\chi_{\rm norm}^{2}$ & 2.4 & 2.7 & 3.2 & 2.1 & 2.3 & 2.2 & 2.6 & 2.9 & 3.6 \\
Fraction within $1\sigma$, $f_{1\sigma}$ & 58\% & 50\% & 58\% & 61\% & 54\% & 68\% & 57\% & 48\% & 54\%\\
Fraction within $2\sigma$, $f_{2\sigma}$ & 90\% & 83\% & 89\% & 88\% & 86\% & 92\% & 90\% & 82\% & 87\%\\
Fraction within $3\sigma$, $f_{3\sigma}$ & 95\% & 95\% & 93\% & 96\% & 96\% & 95\%  & 95\% & 95\% & 93\%\\
Supported-arm fraction, $f_{sup}$ & 0.48 & 0.64 & 0.66 & 0.28 & 0.43 & 0.46 & 0.38 & 0.54 & 0.57\\
Combined score, $S$ & 1.4 & 1.3 & 1.2 & 1.5 & 1.5 & 1.2 & 1.5 & 1.5 & 1.4\\
\enddata
\end{deluxetable*}

Figure~\ref{figs:densitymap} presents the spatial distribution of the Classical Cepheids. The red and blue symbols denote subsamples A and B, respectively. The four panels show the distributions in the Galactic ($X-Y$) plane, the vertical height ($Z$) as a function of Galactocentric radius ($R$), $Z$ as a function of heliocentric distance ($d$), and the $W1$-band extinction ($A_{W1}$) as a function of $d$. The distance from the Sun to the Galactic center ($R_0$) used in this article is 8.15 kpc~\citep{reid2019}.

Both subsamples are predominantly concentrated close to the Galactic mid-plane, as expected for young disk stars. The vertical dispersion appears to increase toward larger Galactocentric and heliocentric distances. Meanwhile, the $d-A_{W1}$ distribution indicates that extinction generally increases with heliocentric distance. The resulting loss of sensitivity to highly reddened sources likely contributes to the decreasing number of observed Cepheids at large distances. This behavior suggests that the observed sample is affected by both interstellar extinction and the distance-dependent detection limit. In particular, distant Cepheids located close to the Galactic mid-plane are more likely to be missed because their lines of sight generally suffer stronger extinction.

Within a heliocentric distance of approximately 10~kpc, the Cepheid
distribution is sufficiently dense and continuous to provide a robust
view of the principal spiral-arm structure. At distances of
$\sim10$--15~kpc, the sample becomes progressively less complete, but
the remaining Cepheids still delineate several coherent distant arm
segments and offer a first glimpse of their possible extensions and
large-scale morphology.

Previous studies have primarily identified spiral-arm structures directly from the spatial distributions of stellar samples \citep{lemasle2022,drimmel2025}. However, Galactic spiral arms are intrinsically non-uniform and contain local features such as spurs and segments \citep{xu2021}. Given that young stars are expected to retain substantial information about their natal spiral arms \citep{xu2021,hao2021,ge2024}, we quantitatively compare the Cepheid distribution with the R19 
and X23 models using the minimum point-to-arm distance, $d_i$, the 
normalized point-to-arm distance, $\delta_i$, and the supported-arm 
fraction, $f_{\rm sup}$ defined in Appendix~\ref{secA1} The former two quantities 
characterize the proximity of the Cepheids to the modeled arms, whereas 
$f_{\rm sup}$ describes the fraction of the modeled arm length covered 
by the current Cepheid sample.

Because this work focuses on the large-scale spiral morphology of the Milky Way, we consider only the major spiral arms and exclude smaller-scale features, such as local spurs. To avoid possible confusion with bar-related inner structures, such as the 3-kpc arms, we do not assign Cepheids to spiral arms within a Galactocentric radius of $R_{\rm GC}<3$~kpc. Our analysis therefore does not attempt to distinguish the 3-kpc arms from the innermost extensions of the major spiral arms in this region.
In this comparison, the R19 model contains five long and relatively
continuous major arms (Scutum--Centaurus--OSC arm, Norma--Outer arm, Perseus arm, Local arm, and Sagittarius--Carina arm) with a total length of $\sim222$~kpc, whereas the
X23 model contains seven major arm structures with a shorter total length
of $\sim173$~kpc.

The two models show somewhat different behavior in the two 
complementary geometric comparisons summarized in Table~\ref{tab:model_comparison}. 
When the distance is measured from each Cepheid to its nearest arm locus, 
R19 yields slightly smaller offsets than X23, indicating a marginally 
closer local point-to-arm correspondence for the current sample. 
Within $R_{\rm GC}\leq15$~kpc, the total modeled arm lengths are 
approximately 222.5~kpc for R19 and 172.8~kpc for X23. A model containing more 
numerous or extended arm loci provides more opportunities for an 
individual Cepheid to lie close to one of the predicted curves and may 
therefore yield smaller point-to-arm distances.

\begin{figure*}[!ht]
    \centering
    \includegraphics[width=0.48\linewidth]{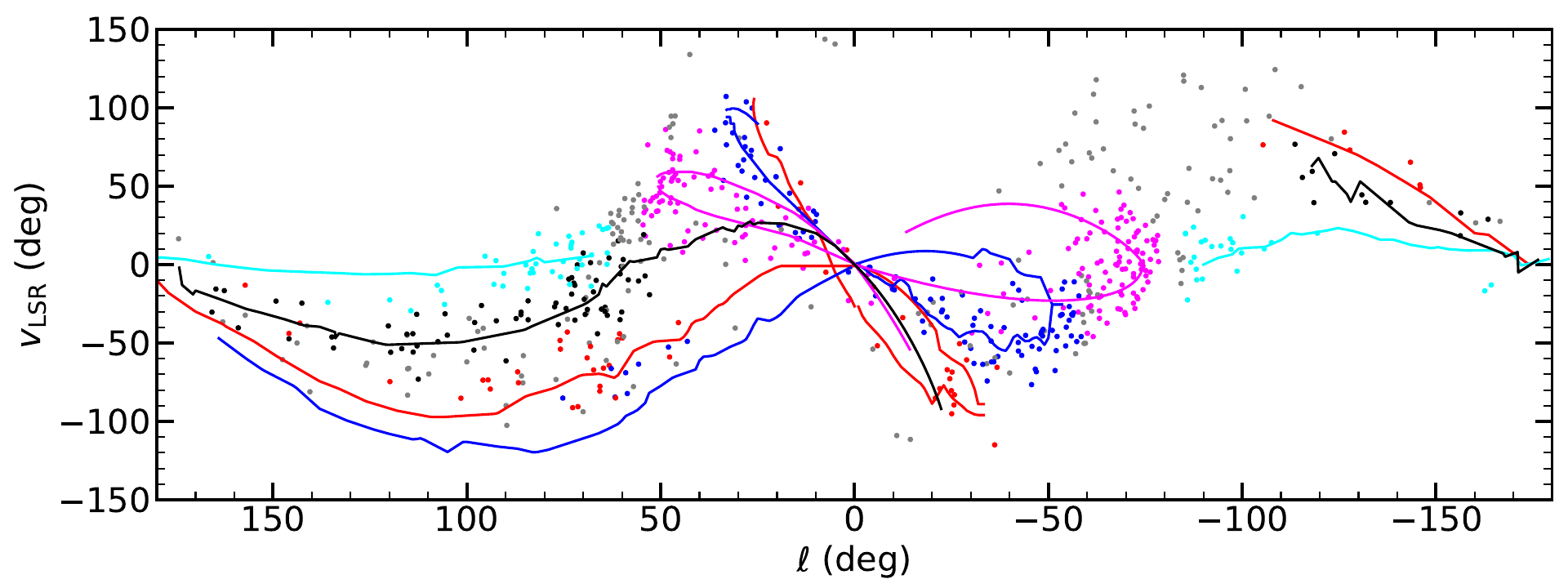}
    \includegraphics[width=0.48\linewidth]{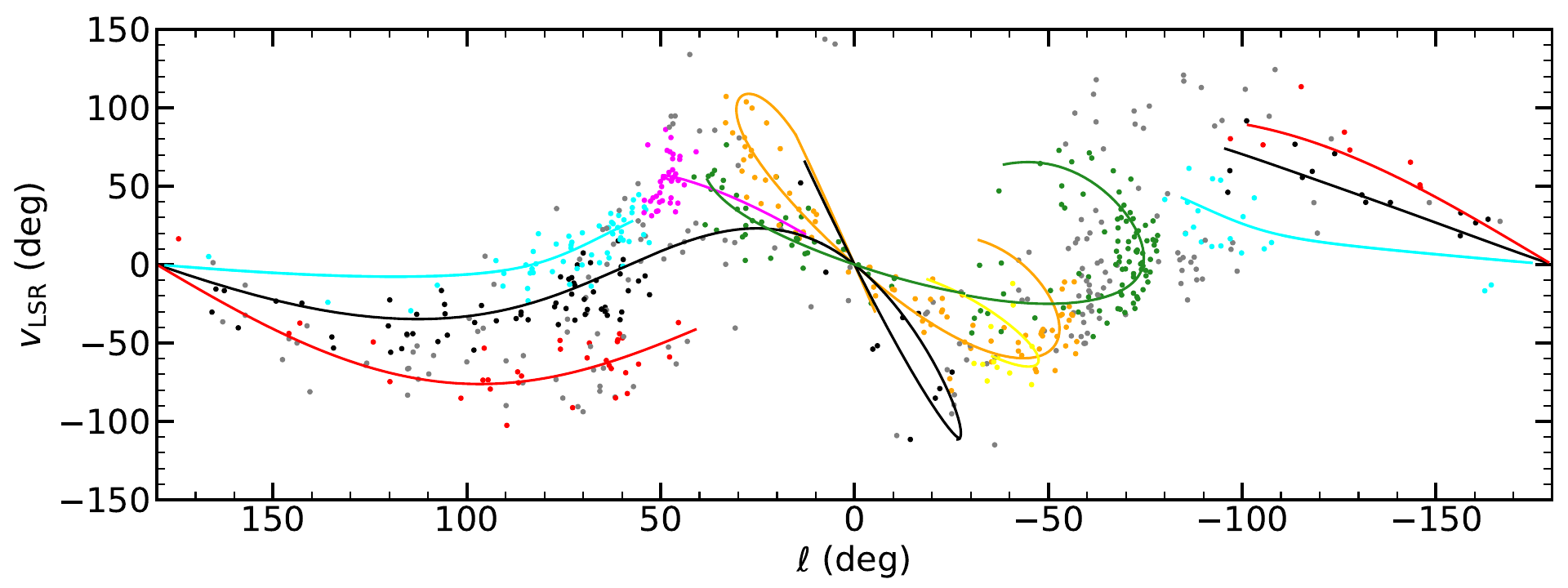}\\
    \includegraphics[width=0.48\textwidth, trim=1cm 2cm 0cm 0cm, clip]{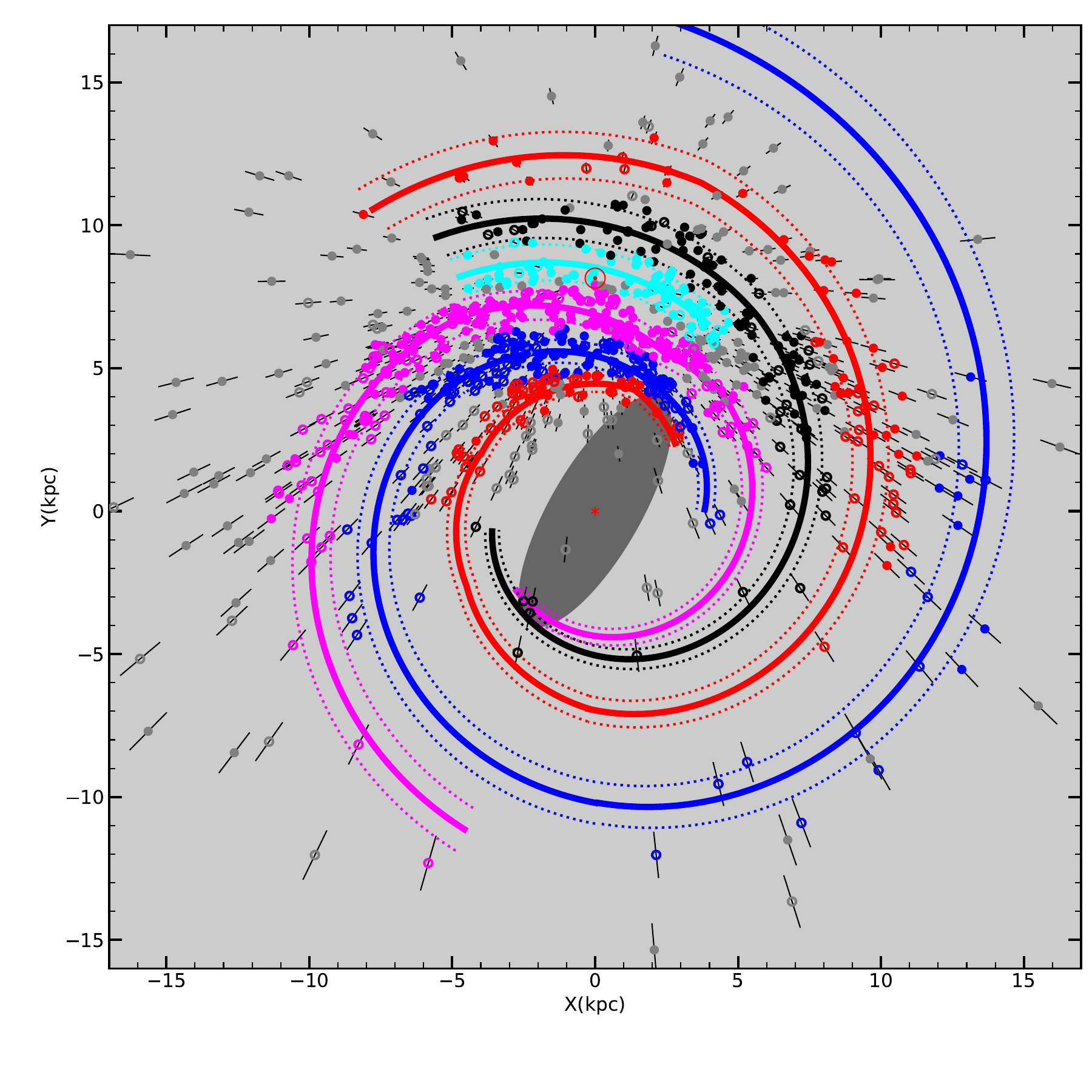}
    \includegraphics[width=0.48\textwidth, trim=1cm 2cm 0cm 0cm, clip]{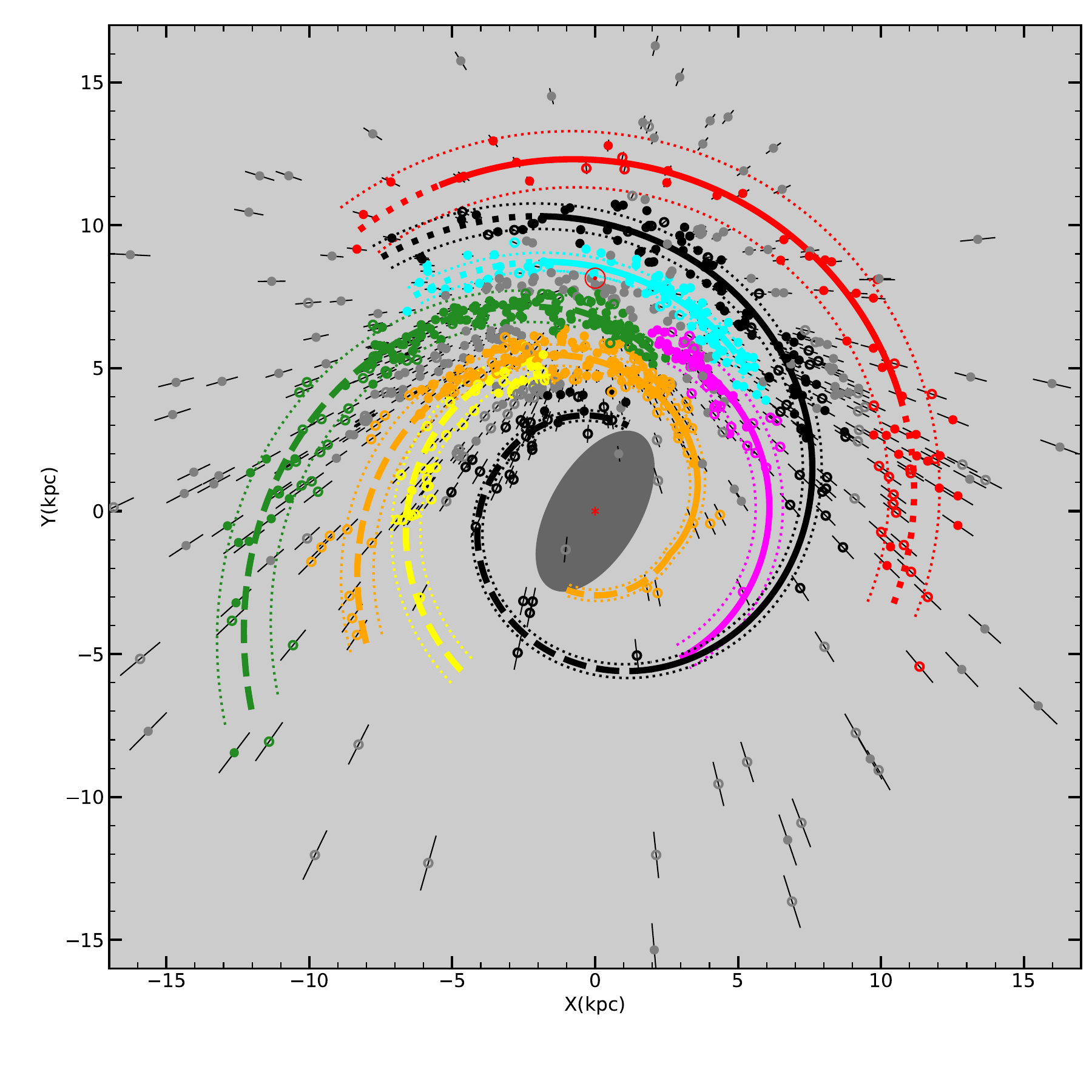}
    \caption{Left: R19 model; right: X23 model. Top: $\ell-v_{\rm LSR}$; bottom: $X-Y$. Points denote classical Cepheids with ages $\leq100$ Myr. Colors indicate spiral arms. R19 (middle panel): Scutum--Centaurus--OSC arm (blue), Norma--Outer arm (red), Perseus arm (black), Local arm (cyan), Sagittarius--Carina arm (magenta). X23 (right panel): Outer arm (red), Perseus arm (black), Local arm (cyan), Sagittarius arm (magenta), Carina arm (dark green), Norma arm (orange), and Centaurus arm (yellow).  Colored circles represent classical Cepheids (filled: with $v_{\rm LSR}$; open: without), with colors indicating assigned spiral arms. For R19, the model $v_{\rm LSR}$ values follow the kinematic parameters of \citet{reid2019}; for X23, they are calculated from the spatial arm loci using the Galactic rotation model adopted by \citet{reid2019}. The bar geometries shown in the two panels follow different literature-based representations: the relatively long bar adopted for the R19 four-arm model is taken from \citet{wegg2015}, whereas the shorter bar shown with the X23 multiple-arm model follows \citet{hilmi2020}, as adopted by X23.}
    \label{figs:assigned_spiral_arm}
\end{figure*}

The supported-arm fraction provides a complementary 
arm-to-source comparison: it measures the fraction of the modeled arm 
length for which at least one Cepheid in the current sample lies within 
0.5~kpc. For the present sample, the supported-arm fractions are 0.44 
for R19 and 0.64 for X23. These values describe the extent to which the 
respective published arm networks are covered by the currently available 
Cepheids. They should not be interpreted as direct evidence against arm 
segments lacking nearby Cepheids, because the measured fractions depend 
on the number and extent of the modeled arms, the adopted support radius, 
and the spatial completeness of the Cepheid catalogue. In particular, distant and highly extincted regions are less 
completely sampled and may contain genuine arm segments without currently 
detected Cepheids. Using near-infrared classical Cepheids, 
\citet{minniti2021} demonstrated that highly obscured regions on the far 
side of the Galactic disk can be probed and provided observational 
constraints on its distant spiral-arm structure. Future infrared surveys 
are expected to identify additional Cepheids in these regions and provide 
a more complete assessment of the distant spiral arms.

\begin{deluxetable*}{ccccc}[!t]
\tablecolumns{5}
\tablecaption{Parameter of Classical Cepheids\label{tab:arm_assigned}}
\tablehead{
\colhead{No.} & \colhead{column}& \colhead{Description} & \colhead{Unit} & \colhead{Example} 
}
\startdata
1 & Name & Source name & ... & J002230.35+644839.6\\
2 & Gaia ID & Gaia source ID & ... & 429385923752386944\\
3 & glon & Galactic Longitude & deg & 119.860\\
4 & glat & Galactic Latitude & deg & +2.110\\
5 & dist & Distance based on AllWISE/unWISE & kpc & 5.96\\
6 & e\_dist &  Uncertainty in distance based on AllWISE/unWISE & kpc & 0.24\\
7 & age &  Age estimate & Myr & 86\\
8 & g\_mag & Gaia G-band magnitude & mag & 10.97\\
9 & A\_w1 & Extinction in W1 based on averaged W1 & mag & 0.19\\
10 & V\_lsr & Local Standard of Rest Velocity & km s$^{-1}$ & -74.64\\
11 & arm\_R19 & Arm allocation based on the R19 model & ... & NOA\\
12 & arm\_X23 & Arm allocation based on the X23 model & ... & Out\\
13 & arm\_new & Arm allocation in Section~\ref{sec3.3} & ... & Out\\
\enddata
\tablecomments{Entries 1--9 (written with different font) are taken from~\citet{skowron2025}, Gaia, and AllWISE catalogs. The remaining entries had been calculated in this work. (This table is available in its entirety in machine-readable form in the online article.)}
\end{deluxetable*}

Overall, both R19 and X23 provide reasonably good descriptions 
of the current Cepheid distribution. Approximately 95\% of the Cepheids 
lie within $3\sigma$ of an arm locus in either model. R19 yields slightly 
smaller local point-to-arm distances, whereas X23 has a higher 
supported-arm fraction for the currently available Cepheid sample. 
The combined scores are $S=1.4$ for R19 and $S=1.3$ for X23. 
The slightly lower value of $S$ for X23 suggests a marginally 
better overall correspondence under the adopted definition, although 
this small difference is based on the currently available sample, whose 
completeness decreases in distant and highly extincted regions. It is therefore insufficient to distinguish decisively between 
the two models.
More complete samples, particularly from future infrared surveys, will 
be required to assess the distant and obscured arm segments more 
reliably.

\subsection{Spiral-Arm Reassignment and Comparison of the Refit Models}

To further examine how well the two models describe the spiral 
structure traced by the Cepheids, we first assign the individual 
Cepheids to the corresponding spiral arms under each model using the 
arm-association probability 
$P_{*}^{\rm arm}(X,Y,V_{\rm LSR})$ defined in 
Appendix~\ref{secB}.

For the subsample with available radial velocities, we examine the spiral-arm associations independently in the $X$--$Y$ plane and the $\ell$--$v_{\rm LSR}$ diagram. Among the Cepheids spatially assigned to spiral arms, application of the additional kinematic criterion retains 490 of the 525 assignments for the R19 model and 459 of the 482 assignments for the X23 model, corresponding to 93.3\% and 95.2\%, respectively. The velocity offsets relative to the assigned arm tracks are also centered near zero, with median values of $0\pm0.6~\mathrm{km s^{-1}}$ for the full sample and $0\pm1.2$ and $0\pm0.6~\mathrm{km s^{-1}}$ for subsamples A and B, respectively. These results show that Cepheids spatially associated with a modeled arm generally also follow the corresponding gas- and maser-defined kinematic locus. They further suggest that young stars with ages $\lesssim100$ Myr may retain a kinematic ``memory'' of their birth environments. A similar case is presented by \citet{craig2025}, who used classical Cepheids to infer H\textsc{I} distances by matching their distributions with H\textsc{I} structures in $\ell-b-v_{\rm LSR}$ space, thereby tracing the H\textsc{I} morphology.

Radial velocities are unavailable for a substantial fraction of the sample. In particular, only 149 of the 301 Cepheids in subsample A have measured radial velocities. Restricting the arm assignment to sources with $v_{\rm LSR}$ measurements would therefore remove approximately half of the youngest tracers and substantially reduce the spatial coverage and statistical power of the sample. Given the high level of agreement between 
the spatial and kinematic criteria in the radial-velocity subsample, we 
determine arm membership for Cepheids without radial-velocity 
measurements using only the spatial component 
$P_{*}^{\rm arm}(X,Y)$, with the velocity term omitted, rather than the full probability.

\begin{table*}[!htp]
    \caption{Spiral Arm Parameters}
    \label{tab:arm_fit}
    \hspace{-1cm}
    \begin{threeparttable}
    \scriptsize
        \begin{tabular}{lcrrrrrc}
        \hline \hline
Spiral Arm & Number & $\beta$ Range & $R_{\rm ref}$  & $\beta_{\rm ref}$  & $\psi$   & Width &   Tracer   \\ 
  &  &  (deg) &  (kpc) &  (deg) &  (deg) &  (kpc) & 
  \\ \hline
  \multicolumn{8}{c}{R19}
  \\ \hline
  Norma$-$Outer &  47 & 4 $\rightarrow$ 409 & 6.23 $\pm$ 0.09 & 241.0 & 9.75 $\pm$ 0.29 & 0.58 & subsample A\\
  Norma$-$Outer &  76 & -38 $\rightarrow$ 387 & 6.27 $\pm$ 0.09 & 241.0 & 9.22 $\pm$ 0.23 & 0.53 & subsample B\\
  Scutum$-$Centaurus$-$OSC &  81 & $-$280 $\rightarrow$ 98 & 6.08 $\pm$ 0.06 & 241.0 & 10.51 $\pm$ 0.30 & 0.64 & subsample A\\
  Scutum$-$Centaurus$-$OSC &  113 & $-$290 $\rightarrow$ 66 & 6.14 $\pm$ 0.05 & 241.0 & 10.24 $\pm$ 0.25 & 0.59 & subsample B\\
  Sagittarius$-$Carina &  62 & $-$118 $\rightarrow$ 76 & 7.50 $\pm$ 0.07 & $-$16.3 & 11.15 $\pm$ 0.58 & 0.57 & subsample A\\
  Sagittarius$-$Carina &  198 & $-$155 $\rightarrow$ 61 & 7.46 $\pm$ 0.03 & $-$16.3 & 13.04 $\pm$ 0.38 & 0.51 & subsample B\\
  Perseus &  26 & $-$24 $\rightarrow$ 219 & 8.50 $\pm$ 0.12 & 44.5 & 11.10 $\pm$ 1.11 & 0.57 & subsample A\\
  Perseus &  85 & $-$23$\rightarrow$ 263 & 8.66 $\pm$ 0.07 & 44.5 & 11.10 $\pm$ 0.69 & 0.60 & subsample B\\
  Local &  17 & $-$18 $\rightarrow$ 35 & 8.09 $\pm$ 0.11 & 12.3 & 14.17 $\pm$ 2.32 & 0.41 & subsample A\\
  Local &  64 & $-$30$\rightarrow$ 36 & 8.28 $\pm$ 0.06 & 12.3 & 7.93 $\pm$ 1.07 & 0.43 & subsample B
  \\ \hline
  \multicolumn{8}{c}{X23}
  \\ \hline
  Norma &  79 & $-$118 $\rightarrow$ 146 & 5.55 $\pm$ 0.05 & $-$7.7 & 14.67 $\pm$ 0.66 & 0.50 & subsample A\\
  Norma &  88 & $-$100 $\rightarrow$ 143 & 5.57 $\pm$ 0.06 & $-$7.7 & 16.00 $\pm$ 0.99 & 0.56 & subsample B\\
  Perseus &  48 & -35 $\rightarrow$ 371 & 6.73 $\pm$ 0.09 & 131.6 & 9.59 $\pm$ 0.60 & 0.60 & subsample A\\
  Perseus &  93 & -43 $\rightarrow$ 376 & 6.69 $\pm$ 0.10 & 131.6 & 10.11 $\pm$ 0.40 & 0.58 & subsample B\\
  Centaurus &  17 & 267 $\rightarrow$ 344 & 5.72 $\pm$ 0.11 & 315.5 & 7.77 $\pm$ 2.18 & 0.40 & subsample A\\
  Centaurus &  25 & 243 $\rightarrow$ 344 & 5.52 $\pm$ 0.07 & 315.5 & 5.88 $\pm$ 1.63 & 0.41 & subsample B\\
  Sagittarius &  23 & 20 $\rightarrow$ 89 & 6.21 $\pm$ 0.10 & 46.5 & 2.89 $\pm$ 3.08 & 0.45 & subsample A\\
  Sagittarius &  36 & 17 $\rightarrow$ 119 & 6.18 $\pm$ 0.08 & 46.5 & 3.83 $\pm$ 2.34 & 0.33 & subsample B\\
  Carina &  40 & $-$95 $\rightarrow$ 22 & 8.11 $\pm$ 0.06 & $-$29.4 & 18.11 $\pm$ 0.68 & 0.44 & subsample A\\
  Carina &  135 & $-$124 $\rightarrow$ 25 & 7.95 $\pm$ 0.04 & $-$29.4 & 18.32 $\pm$ 0.42 & 0.50 & subsample B\\
  Local &  25 & $-$38 $\rightarrow$ 57 & 7.89 $\pm$ 0.08 & 20.8 & 13.53 $\pm$ 1.27 & 0.38 & subsample A\\
  Local &  64 & $-$42 $\rightarrow$ 55 & 7.97 $\pm$ 0.05 & 20.8 & 11.33 $\pm$ 0.70 & 0.37 & subsample B\\
  Outer &  14 & 4 $\rightarrow$ 105 & 11.65 $\pm$ 0.20 & 57.0 & 3.57 $\pm$ 1.89 & 0.62 & subsample A \\
  Outer &  45 & $-$41 $\rightarrow$ 116 & 11.49 $\pm$ 0.12 & 57.0 & 3.60 $\pm$ 0.66 & 0.77 & subsample B\\
  \hline
  \multicolumn{8}{c}{New}
  \\ \hline
  Norma &  77 & $-$118 $\rightarrow$ 146 & 5.56 $\pm$ 0.04 & $-$2.4 & 14.86 $\pm$ 0.53 & 0.36 & subsample A\\
  Perseus &  53 & -35 $\rightarrow$ 374 & 6.06 $\pm$ 0.10 & 165.8 & 9.48 $\pm$ 0.50 & 0.63 & subsample A\\
  Centaurus &  24 & $-$93 $\rightarrow$ 1 & 5.31 $\pm$ 0.04 & $-$41.4 & 12.52 $\pm$ 0.17 & 0.17 & subsample A\\
  Sagittarius &  20 & 20 $\rightarrow$ 89 & 6.17 $\pm$ 0.10 & 49.3 & 2.00 $\pm$ 3.09 & 0.44 & subsample A\\
  Carina &  40 & $-$95 $\rightarrow$ 22 & 8.10 $\pm$ 0.06 & $-$29.1 & 18.09 $\pm$ 0.68 & 0.44 & subsample A\\
  Local &  24 & $-$38 $\rightarrow$ 57 & 7.77 $\pm$ 0.08 & 24.5 & 13.56 $\pm$ 1.24 & 0.38 & subsample A\\
  Outer &  12 & 4 $\rightarrow$ 101 & 11.27 $\pm$ 0.25 & 63.2 & 5.33 $\pm$ 2.65 & 0.69 & subsample A 
  \\ \hline
  \end{tabular}
  \end{threeparttable}
\end{table*}

For the R19 and X23 models, the distributions of classical Cepheids in the $\ell-v_{\rm LSR}$ diagram and the $X-Y$ plane are shown as dots in Figure~\ref{figs:assigned_spiral_arm}.
Using the R19 and X23 spiral-arm models, 793 and 732 Cepheids, respectively, are assigned to the corresponding arm structures (see Table~\ref{tab:arm_assigned}). We also note that the R19 and X23 models cover different regions and extents of the Galactic disk. Therefore, the number of Cepheids assigned to each model cannot be used directly as a criterion for determining which model provides a better representation of the spiral structure.

\begin{figure*}[!ht]
    \centering
    \includegraphics[width=0.45\textwidth, trim=1cm 3cm 0cm 0cm, clip]{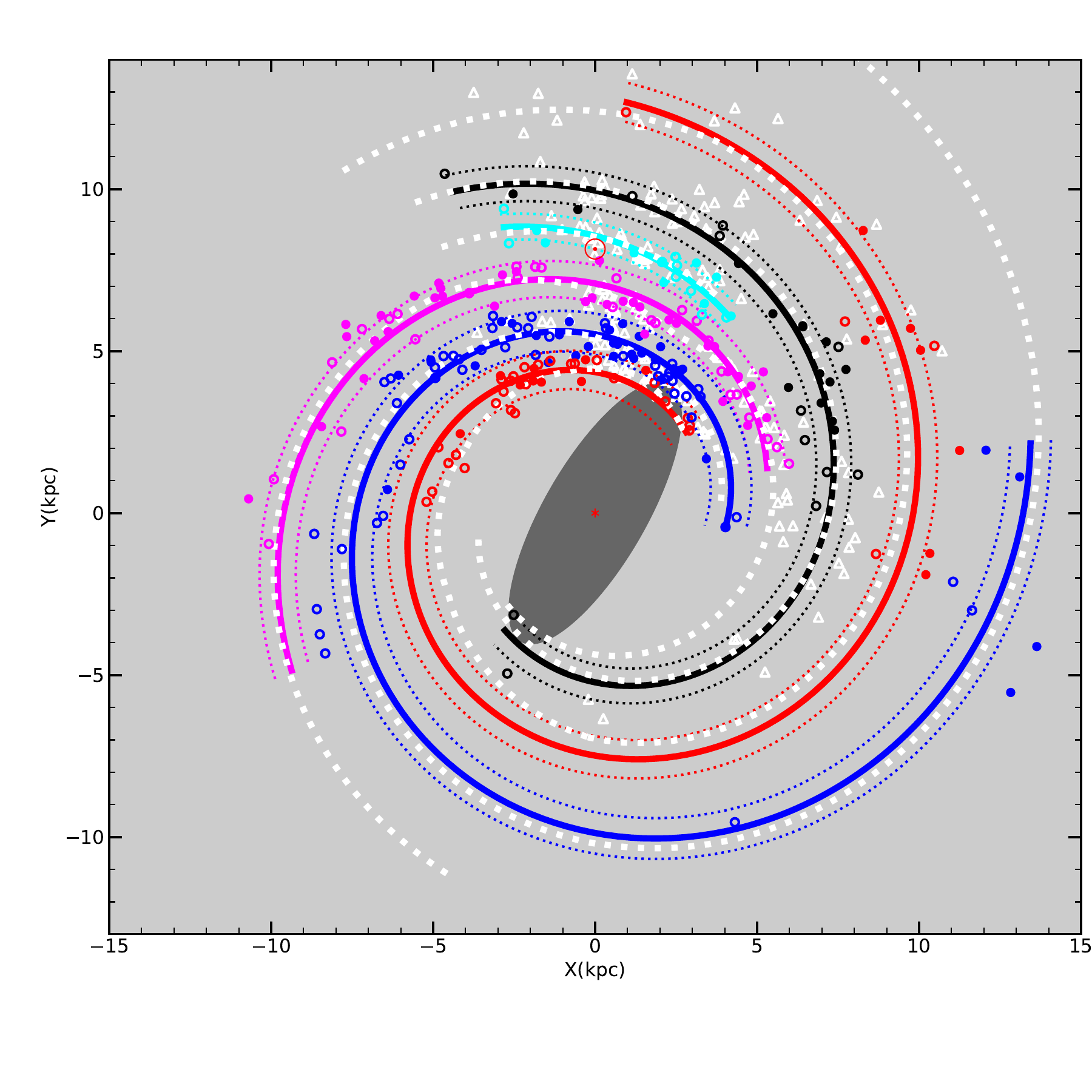}
    \includegraphics[width=0.45\textwidth, trim=1cm 3cm 0cm 0cm, clip]{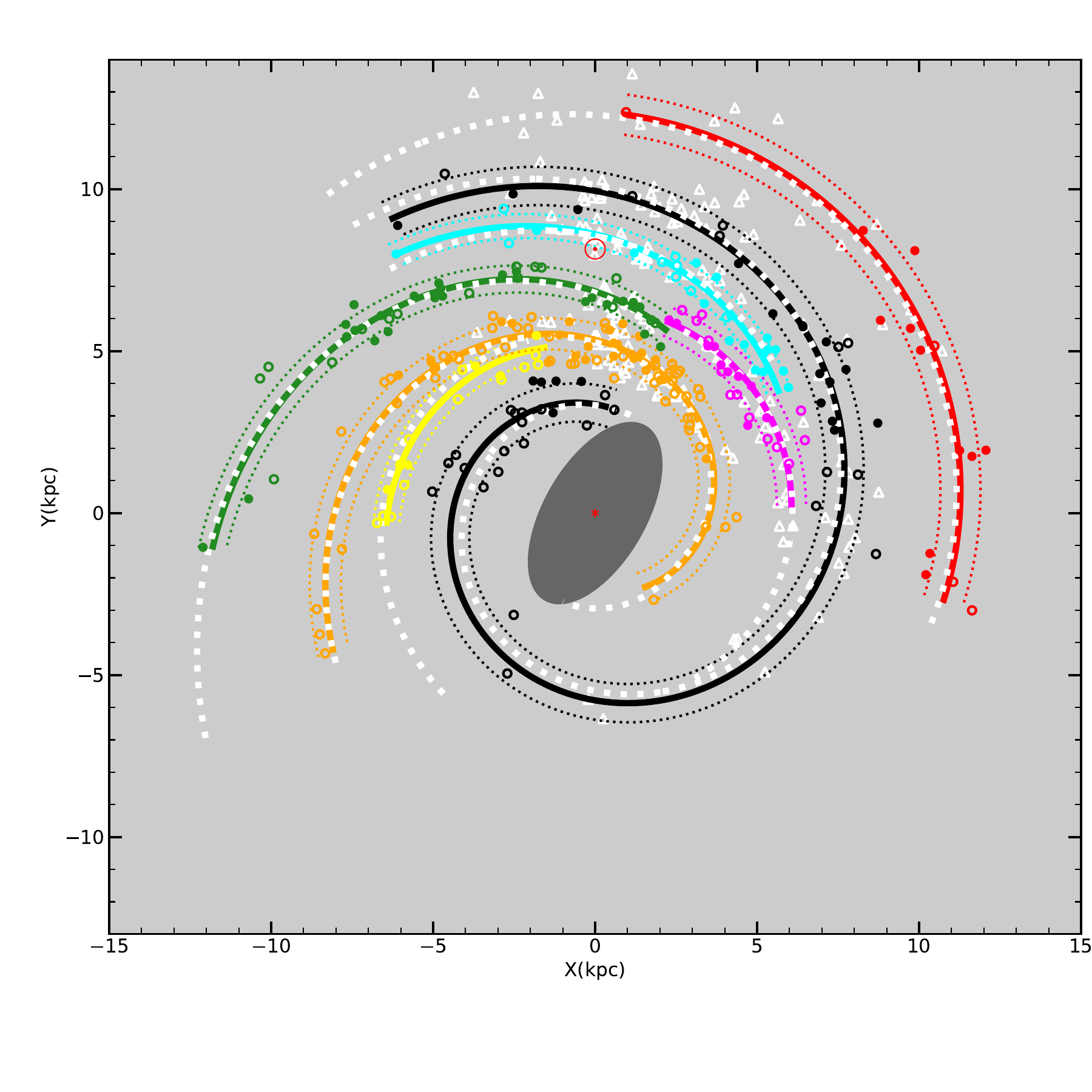}
    \includegraphics[width=0.45\textwidth, trim=1cm 3cm 0cm 0cm, clip]{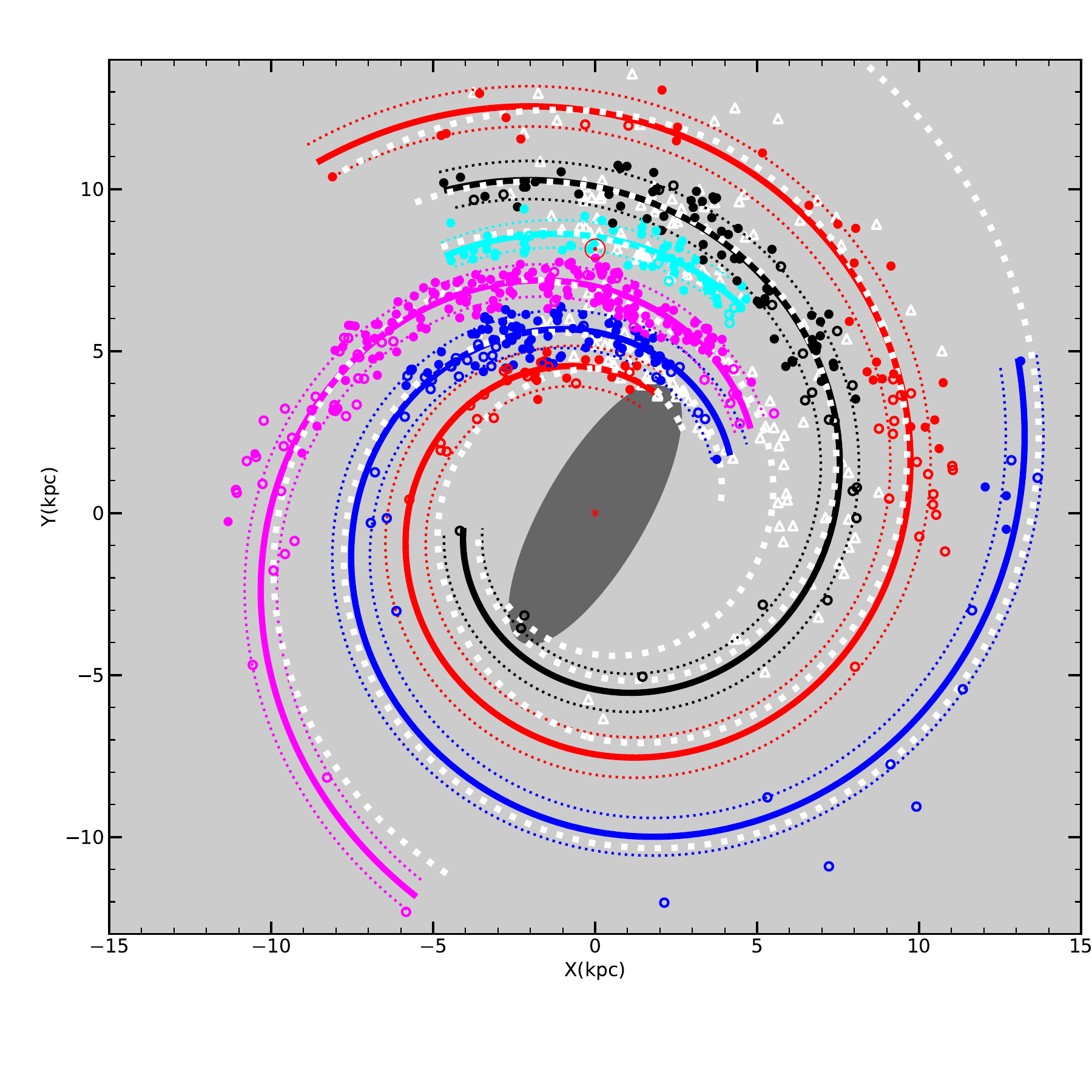}
    \includegraphics[width=0.45\textwidth, trim=1cm 3cm 0cm 0cm, clip]{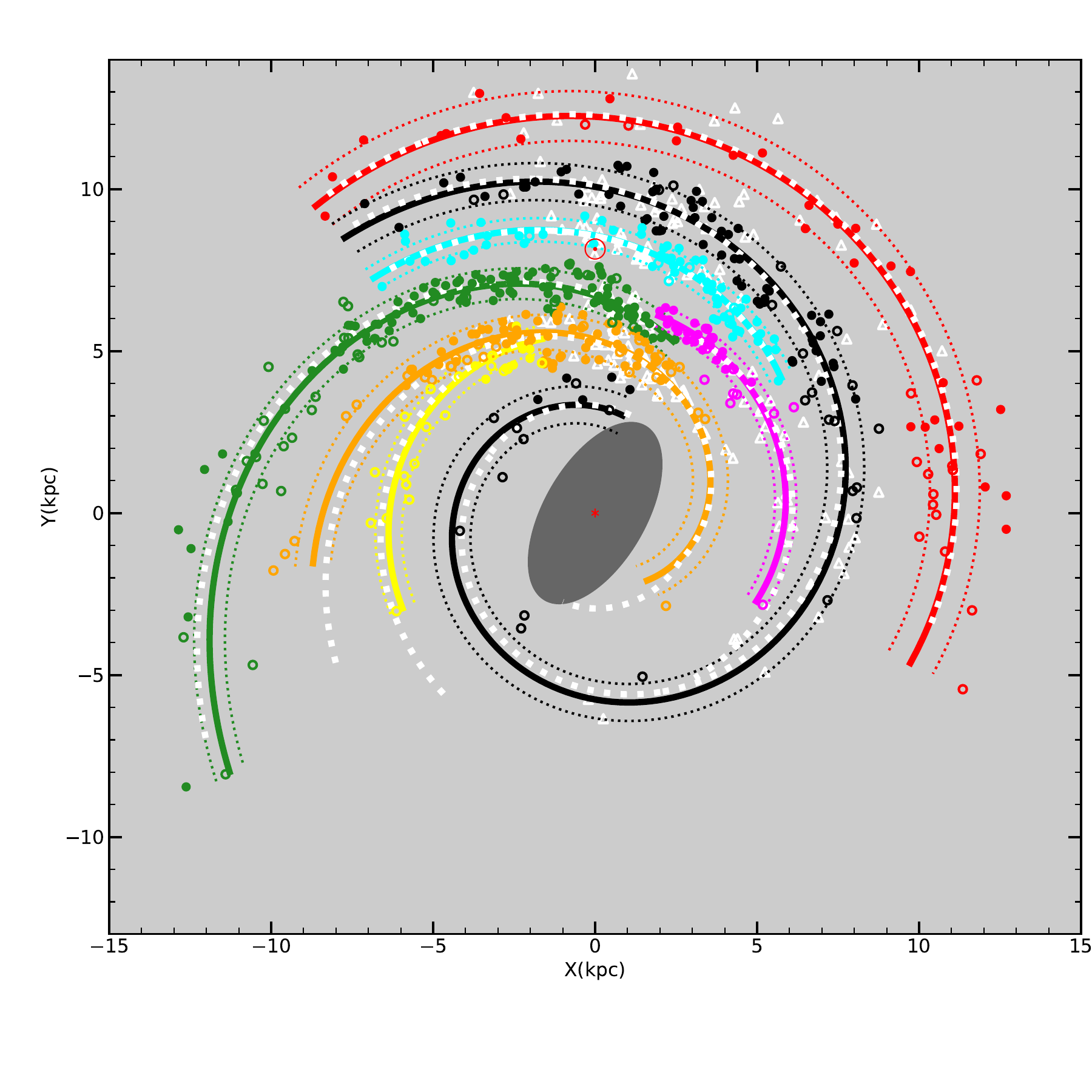}
    \caption{Left: Spiral arm patterns traced by Classical Cepheids of different ages under the four-arm model. Right: Spiral arm patterns traced by Classical Cepheids of different ages under the multiple-arm model. Top: subsample A; bottom: subsample B. White dashed lines show maser-traced spiral arms; white triangles denote masers. Colored circles represent classical Cepheids (filled: with $v_{\rm LSR}$; open: without), with colors indicating assigned spiral arms (see Figure~\ref{figs:assigned_spiral_arm}). The solid lines represent the best-fitting spiral arms traced by Classical Cepheids, with dashed lines indicating their widths.}
    \label{figs:spiral_arm_models}
\end{figure*}

After assigning the Cepheids under the two model frameworks, we  refit the corresponding spiral-arm loci using the assigned 
samples, following the procedure described in 
Appendix~\ref{secC}.
The spiral-arm fitting results traced by young classical Cepheids are summarized in Table~\ref{tab:arm_fit}. The width of each spiral arm is estimated from the dispersion of the tracer positions relative to the arm center. We also examined the dependence of the fitted spiral-arm parameters on the adopted Sun--Galactic Centre distance. Repeating the fitting with $R_0$=8.2~kpc \citep{blandhawthorn2016} and $R_0$=8.0~kpc~\citep{ghez2008} produces only a slight inward contraction of the fitted arm loci in Galactocentric radius. The resulting pitch angles remain consistent with those obtained using the fiducial value of $R_0$ within their uncertainties, indicating that the inferred large-scale spiral morphology is only weakly sensitive to the adopted Sun--Galactic Centre distance.

Figure~\ref{figs:spiral_arm_models} shows the four-arm (left) and multiple-arm (right) spiral structures traced by young Classical Cepheids in different age subsamples. 
Although the spiral structures traced by subsamples A and B show slight
offsets from the maser-defined arms, as indicated by the differences
between the white dashed maser-defined loci and the colored solid
Cepheid-fitted loci, these deviations are generally small. For subsample A, the median offsets of the individual arms range from 0.10 to 0.43~kpc in the R19 model, with a median value of 0.18~kpc over all arms, and from 0.02 to 0.25~kpc in the X23 model, with a median value of 0.12~kpc. For subsample B, the corresponding ranges are 0.15--0.35~kpc and 0.03--0.18~kpc, with median values of 0.25 and 0.16~kpc for the R19 and X23 models, respectively.

The largest offset for subsample A in the R19 model is associated with the Norma--Outer Arm, which contains several kinks. Moreover, the relatively large discrepancies occur mainly in the southern Galactic region, where the maser sampling is sparse. This large offset may therefore partly reflect uncertainties in the adopted maser-defined arm locus rather than a genuine physical displacement of the Cepheid-traced structure.

These small offsets are broadly consistent with the results of \citet{ge2024}, who found that the density peaks traced by B3--B5 and O--B2 stars in different azimuthal directions differed from the nearby maser-based spiral arm by no more than 0.11~kpc. This offset is not significant compared with the typical Galactic spiral-arm width of approximately 0.3~kpc. Observationally, these results suggest that young stars formed in spiral-arm star-forming regions can retain a spatial association with their natal arm structures over timescales of up to $\sim$100~Myr. Within the measurement and fitting uncertainties, the large-scale spiral structures traced by the two Cepheid subsamples are therefore mutually consistent, although the younger subsample remains more closely aligned with the maser-defined arms.

\subsection{An Updated View of the Large-scale Spiral Structure Favored by Young Classical Cepheids}
\label{sec3.3}

Given this, and to provide a clearer view of the spiral structure, we focus on the younger Cepheid subsample (i.e., 20--60~Myr) in this subsection. In the northern sky, masers already provide a well-defined tracing of the arms, and the Cepheid distribution shows a high degree of consistency with these tracers. More importantly, classical Cepheids also trace spiral structures in the southern sky that are currently missing in maser observations, filling existing observational gaps.

The R19 model adopts an inner four-arm spiral morphology, consisting of the Norma, Scutum, Perseus, and Sagittarius arms, whereas the X23 model argues for only two major inner spiral arms, namely the Norma and Perseus arms. In the first Galactic quadrant, X23 noted that the maser sources assigned separately by R19 to the Norma and Scutum arms are intermixed in their spatial distribution and therefore assigned them to a single arm, which they designated as the Norma Arm. On the far side of the first Galactic quadrant, X23 further proposed
that the Perseus and Sagittarius arms converge and connect. Beyond
this connection, the Perseus Arm continues toward the inner Galaxy,
an inward extension that has also been suggested by
\citet{minniti2021} and \citet{hyland2026}. The inner Norma and Perseus Arms constitute an important observational basis for the inner two-arm morphology proposed by X23. In addition, the inner two-arm pattern in the X23 model naturally extends from the two ends of the Galactic bar, while in the R19 model the Galactic bar appears to cross the spiral arms, resulting in a more complex morphology.

\begin{figure}[!ht]
    \centering
    \includegraphics[width=0.3\textwidth]{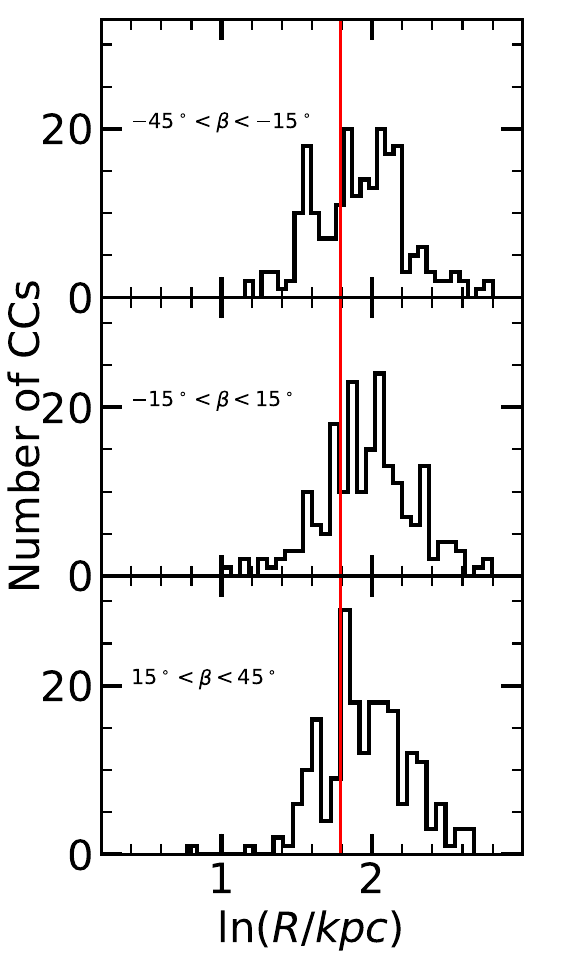}
    \caption{Number density distributions of young Classical Cepheids along different $\beta$ directions. The red vertical line represents 6 kpc.}
    \label{figs:histogram}
\end{figure}

In the left panel of Figure~\ref{figs:spiral_arm_models} (four-arm model), the young Classical Cepheids tracing the Norma arm (red) show an outward offset relative to the maser-based model in the fourth Galactic quadrant, while in the near side of the first quadrant they become mixed with the Scutum arm (blue), favoring a connection between these two arms near $\beta\sim0^\circ$. To examine this possible connection more quantitatively, we calculated the radial number distributions of the young Classical Cepheids in different Galactic-azimuth intervals. In the middle panel of Figure~\ref{figs:histogram}, corresponding to $-15^\circ<\beta<15^\circ$, two local density peaks are present within $R<6\,\mathrm{kpc}$, close to the expected radii of the Scutum and Norma arms in the R19 model. By contrast, in the lower panel, corresponding to $15^\circ<\beta<45^\circ$, the inner distribution is dominated by a single peak. The two arm-associated overdensities therefore become less distinguishable toward positive Galactic azimuths, consistent with a possible connection between the Scutum and Norma arms. In constructing the R19 model, \citet{reid2019} introduced ``kinks'' near the possible intersection region between the Norma and Scutum arms in order to separate the two structures. However, \citet{xu2023} pointed out that the Norma and Scutum arms are fully mixed in both the $\ell-v_{\rm LSR}$ and $X-Y$ distributions in the R19 model, suggesting that they may actually belong to the same spiral arm. Moreover, the maser distribution near $\ell = 0^\circ$ shows a marked broadening, further suggesting the presence of arm bifurcation.

\begin{figure*}[!ht]
    \centering
    \includegraphics[width=0.9\linewidth, trim=1cm 3cm 0cm 0cm, clip]{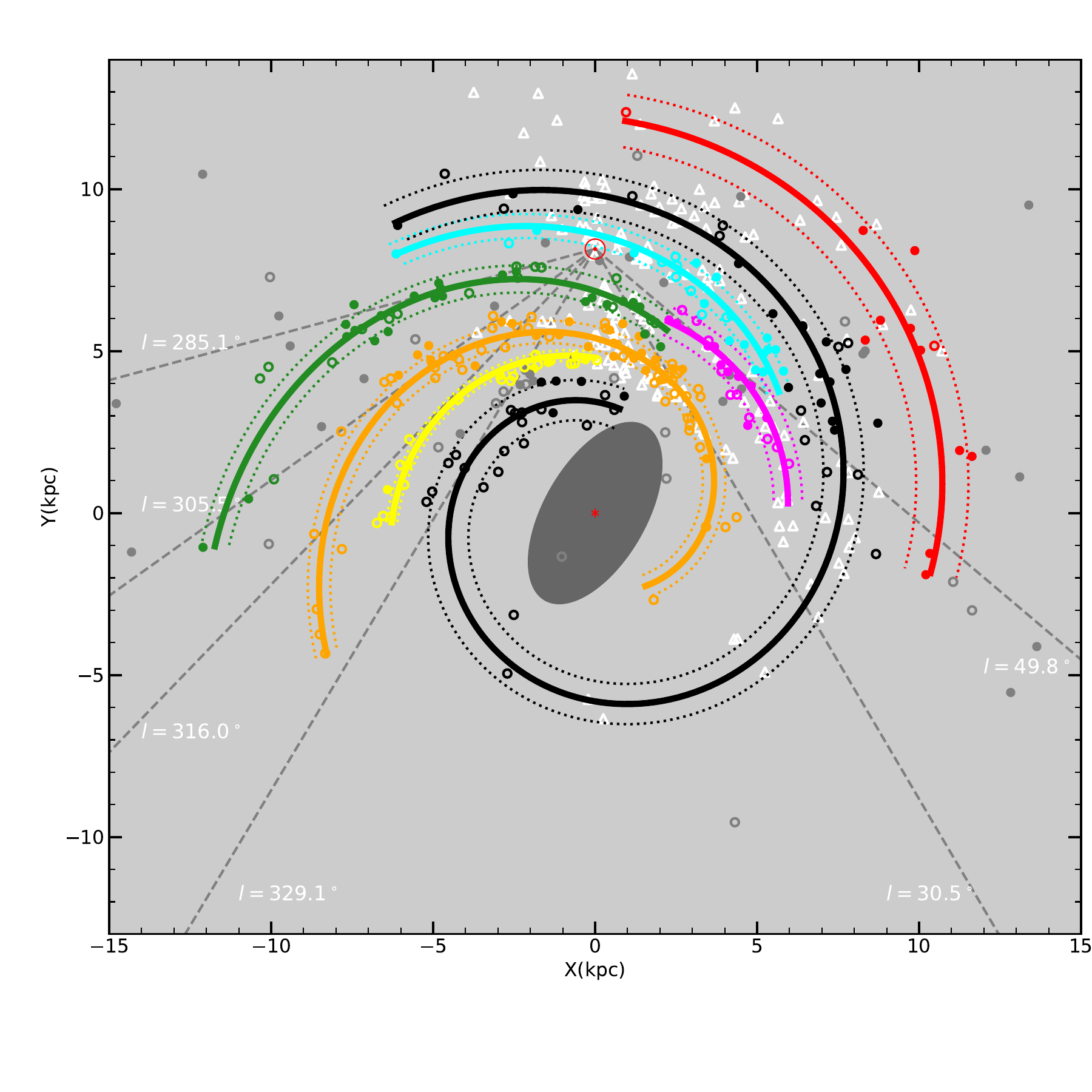}
    \caption{Large-scale spiral structure traced by 20--60~Myr classical Cepheids. Colored circles represent classical Cepheids (filled: with $v_{\rm LSR}$; open: without), with colors indicating assigned spiral arms (see Figure~\ref{figs:assigned_spiral_arm}).White triangles denote masers.}
    \label{figs:general_view}
\end{figure*}

Although the young classical Cepheid lacks coverage in the distant regions of the Sagittarius and Perseus Arms in the first quadrant, recent maser-based studies suggest a possible merging of these two arms \citep{bian2024,hyland2026}. With the inclusion of young classical Cepheids in the southern sky and the maser-indicated convergence of the distant Sagittarius and Perseus arms, the Perseus and Norma arms emerge as the two major inner spiral arms of the Milky Way. This result further supports the ``inner two-arm and outer multi-arm'' spiral morphology proposed by the X23 model~\citep[see also][]{minniti2021}. Statistical studies of external galaxies suggest that such a spiral pattern accounts for approximately 74\% of spiral galaxies and is considered to be a more common type of spiral structure \citep{elmegreen1989, wei2024}. There is no barred spiral galaxy in nearby spiral galaxies with four continuous spiral arms from the inner to the outer regions \citep{wei2024}. Therefore Combining X23 spiral-arm models and the Cepheid data, we construct an updated multiple-arm model, as shown in Figure~\ref{figs:general_view}.

It should be noted that the Centaurus Arm in the X23 model is not directly delineated by maser parallaxes; instead, its initial locus was inferred primarily from the arm-tangent direction near $l\sim311.5^\circ$. In the X23 framework, the Centaurus Arm approaches and eventually connects with the Norma Arm in the fourth Galactic quadrant. The new Cepheid distribution provides additional information on this geometry. In particular, a substantial number of Cepheids are found between the adopted Norma and Perseus loci and extend toward Galactic longitude $l\sim0^\circ$. This distribution suggests that, if these stars trace the continuation of the Centaurus Arm, the Centaurus--Norma junction may lie closer to $l\sim0^\circ$ than indicated by the original X23 locus.
The revised Cepheid assignments significantly affect the fitted
Centaurus Arm. After reassigning the Cepheids according to the
updated arm geometry, the two arms become more clearly separated in
this region. Consequently, the dispersion of the associated Cepheids
around the fitted arm centers is reduced, resulting in a smaller
inferred arm width. The new model also indicates that the Centaurus
Arm does not connect with the Norma Arm in the fourth Galactic
quadrant, leading to a substantial change in the fitted pitch angle
relative to the X23 model.

In barred spiral galaxies, the major spiral arms are commonly observed to originate from the two ends of the central bar \citep{kormendy2004,buta2013,dobbs2014}. To avoid the Galactic bar crossing the inner spiral arms, we follow the X23 model and adopt the shorter bar proposed by \citet{hilmi2020}.
In most multiple-arm spiral galaxies, the spiral structure commonly exhibits symmetric bifurcations or broadenings of the inner spiral arms \citep{elmegreen1989,wei2024}. In the new model, the Norma arm broadens near $\beta\sim0^\circ$ and branches into the Centaurus arm, and the Perseus arm branches into the Sagittarius arm at $\beta \approx 180^{\circ}$. This indicates that the Milky Way spiral structure likely exhibits symmetric bifurcations, resembling the spiral morphologies commonly observed in external multiple-arm spiral galaxies.

Tangent directions provide an independent and important constraint for testing spiral arm models. In Figure~\ref{figs:general_view}, gray dashed lines mark six possible tangent directions reported in the literature \citep{hou2015, kong2025}. 
The new model shows agreement with these tangent directions, despite the fact that the tangent constraints were not used as input parameters in constructing the model.
In the first quadrant, two tangent directions ($\ell = 30.5^\circ$ and $49.8^\circ$) align well with the corresponding positions in the spiral arm models. Thanks to the young classical Cepheid, in the fourth quadrant the model also shows good agreement with four possible tangent directions. We also note that the two nearby spiral-arm tangent directions in the
fourth Galactic quadrant are located at approximately $305.5^\circ$
and $310.4^\circ$, with the latter having an uncertainty of
$\sim5^\circ$~\citep{kong2025}. Figure~6 of \citet{kong2025} presents the radial-velocity
distributions of Hi-GAL clumps over the longitude range
$300^\circ<l<340^\circ$. In the $308^\circ<l<316^\circ$
interval, the component associated with the Centaurus Arm is
relatively weak and difficult to distinguish, whereas it becomes
prominent in the adjacent $316^\circ<l<324^\circ$ interval. This
change in prominence suggests that the tangent direction of the
Centaurus Arm may lie near the boundary of the two intervals, at
approximately $l\sim316^\circ$, although the $8^\circ$ bin size
precludes a precise determination.

We treat Sagittarius and Carina as separate fitted arms 
following the arm segmentation adopted by X23. This separation is not 
based on a clearly detected discontinuity in either the $X$--$Y$ or 
$\ell$--$v_{\rm LSR}$ distribution. Instead, it reflects the substantial 
change in geometry between the two segments. For the youngest Cepheid 
sample in our updated model, the fitted pitch angles are 
$2.00\pm3.09^\circ$ for Sagittarius and $18.09\pm0.68^\circ$ for Carina. 
A single continuous logarithmic spiral would therefore require a 
pronounced kink or a substantial change in pitch angle to connect the 
two segments. Although R19 refers to them jointly as the 
Sagittarius--Carina Arm, its piecewise arm prescription likewise 
requires a kink and a large pitch-angle change between the Sagittarius 
and Carina portions. We therefore follow X23 and list the two segments 
separately for the purpose of arm assignment and fitting. This 
parametrization does not exclude the possibility that they are 
physically connected through a transition region. 

Due to the lack of observational tracers in the distant regions, the connections of the Outer and OSC arms with those in the fourth quadrant remain uncertain. The R19 model proposes possible connections of these arms with the fourth-quadrant spiral structure, but their validity requires further observational confirmation. We also evaluate the updated model using the same comparison 
procedure described in Appendix~\ref{secA1}. As shown in Table~\ref{tab:model_comparison}, it generally provides 
an improved correspondence with the Cepheid data, with smaller median 
point-to-arm distances and lower combined scores.

Although both \citet{drimmel2025} and the present study use the
\citet{skowron2025} catalogue, they differ in their sample-selection
criteria and analysis strategies. \citet{drimmel2025} selected 2\,857 dynamically young
Cepheids by requiring their ages to be shorter than the local
epicyclic period and identified spiral-arm overdensities without
pre-assigning individual sources to specific arms. Because this
age limit increases with Galactocentric radius, their sample
includes many older outer-disk Cepheids that are excluded by our
fixed age limit of 100~Myr. Our analysis instead focuses on the
very young population and examines its spatial and kinematic
association with the maser-defined R19 and X23 arm frameworks.
The two samples overlap substantially within
$R_{\rm GC}\lesssim6$--7~kpc, whereas their difference becomes
increasingly important in the outer disk. The differences in the resulting spiral-arm models can therefore be attributed, at least in part, to the different sample selections, arm-identification methods, and scientific objectives adopted in the two studies.

Within a heliocentric distance of $\sim10$~kpc, the young-Cepheid
sample remains sufficiently dense and continuous to provide reliable
constraints on the principal spiral arms. Owing to the radial age
gradient, however, the fraction of Cepheids younger than 100~Myr
decreases markedly beyond $R_{\rm GC}\sim10$~kpc, particularly for
the 20--60~Myr subsample, resulting in progressively sparser coverage
of the outer disk. Nevertheless, Cepheids at heliocentric distances
of $\sim10$--15~kpc still delineate several distant arm segments and
provide a preliminary view of their possible extensions.

\section{summary}
\label{sec5}

In summary, the young classical Cepheid sample delineates the 
large-scale spiral structure of the Milky Way and shows good overall 
correspondence with the maser-defined spiral arms. In both the R19 and 
X23 models, approximately 95\% of the Cepheids lie within $3\sigma$ of 
at least one arm locus. The two models show different strengths in the 
descriptive statistics: R19 yields slightly smaller local point-to-arm 
distances, whereas X23 has a higher supported-arm fraction. Their 
combined scores of $S=1.4$ for R19 and $S=1.3$ for X23 indicate broadly 
comparable overall correspondence, and the present Cepheid sample does 
not provide a decisive preference between the two global spiral 
morphologies. 

Comparisons between the Cepheid-fitted and maser-defined 
arms further show that the younger subsample (20--60~Myr) has slightly 
smaller typical offsets than the older subsample (60--100~Myr). Even 
for Cepheids approaching $\sim100$~Myr, these offsets generally remain 
smaller than the widths of the spiral arms.

Within the model-dependent arm-assignment framework adopted here,
the distribution of the youngest Classical Cepheids (20--60~Myr)
is consistent with an X23-like spiral morphology comprising two dominant inner 
arms and several additional arm segments. Relative to the morphology 
proposed by \citet{xu2023}, the Cepheid data provide additional 
constraints on several southern and distant spiral-arm segments, some 
of which extend to heliocentric distances of $\sim15$~kpc.

In the future, parallax measurements of distant masers with facilities such as the VLBA, SKA-VLBI, and ngVLA are expected to further reveal the spiral structure in the Galaxy. Meanwhile, as the parallax precision of the \textit{Gaia} mission continues to improve, it is anticipated that by 2030 the accuracy may reach $\sim 7~\mu$as, enabling the resolution of fine spiral arm structures within $\sim 15$~kpc of the Sun.

\begin{acknowledgments}
We thank the anonymous referee for the useful suggestions and comments.
This work was funded by the National SKA Program of China (Grant No. 2022SKA0120103), the National Key R\&D Program of China (grant No.2024YFA1611504), the NSFC Grands 12403077, 12403041, 12503071, 11933011 the National SKA Program of China (grant No. 2022SKA0120103), the Xinjiang Talent Development Fund (No. XJRC-2025-KJ-YJ-CXPT-180), and the Tianshan Talent Training Program of Xinjiang Uygur Autonomous Region under grant No. 2024TSYCTD0013, 2022TSYCLJ0005, and the Key Laboratory for Radio Astronomy. 
This work has made use of data from the European Space Agency (ESA) mission \emph{Gaia} (\url{www.cosmos.esa.int/gaia}), processed by the Gaia Data Processing and Analysis Consortium (DPAC, \url{www.cosmos.esa.int/web/gaia/dpac/consortium}). 
\end{acknowledgments}

\appendix

\section{Quantitative comparison of the spiral-arm models}
\label{secA1}

All statistics defined in this appendix were calculated using the complete parent sample before any model-dependent arm assignment or rejection. Specifically, the full-sample statistics use all 985 Cepheids younger than 100 Myr, while the age-dependent statistics use all 301 and 684 Cepheids in subsamples A and B, respectively.

The spiral-arm loci were
represented by a set of ordered points connected by finite line
segments in the Galactic plane. For the $i$th Cepheid at position
($\boldsymbol{x}_i$, $\boldsymbol{y}_i$), we calculated its minimum distance to all of the
arm segments as
\begin{equation}
    d_i =
    \min_j \min_s
    \left\|
        (\boldsymbol{x}_i,\boldsymbol{y}_i)-\boldsymbol{C}_j(s)
    \right\|,
    \label{eq:point_arm_distance}
\end{equation}
where $\boldsymbol{C}_j(s)$ represents the position along the $j$th
arm segment. Thus, each source was associated with the arm yielding the
smallest projected distance. 

To account for both the positional uncertainty of each source and the
finite physical width of a spiral arm, we further defined the
normalized point-to-arm distance
\begin{equation}
    \delta_i =
    \frac{d_i}
    {\sqrt{\sigma_i^2+w_j^2}},
    \label{eq:normalized_distance}
\end{equation}
where $\sigma_i$ is the adopted positional uncertainty of the $i$th
Cepheid in the Galactic plane and $w_j$ is the width of its
nearest arm. The latter accounts for the fact that young tracers are
distributed across a finite region around the arm center line rather
than exactly along a one-dimensional locus. 

The typical local agreement was characterized using the median and
root-mean-square values of $d_i$, the median value of $\delta_i$, and
the mean squared normalized distance
\begin{equation}
    \chi_{\rm norm}^{2}
    =
    \frac{1}{N}
    \sum_{i=1}^{N}\delta_i^2.
    \label{eq:normalized_chi2}
\end{equation}

Because the two spiral-arm models were fixed independently of the
present Cepheid sample, no fitted parameters were subtracted from
$N$. We also calculated the source-coverage fraction
\begin{equation}
    f_{n\sigma}
    =
    \frac{1}{N}
    \sum_{i=1}^{N}
    I\left(\delta_i\leq n\right),
    \label{eq:source_coverage}
\end{equation}
for $n=1$, 2, and 3, where $I$ is the indicator function. These
statistics measure how closely the observed sources follow at least
one of the predicted arm loci. Since the normalized residuals are not
necessarily independent and Gaussian distributed,
$\chi_{\rm norm}^{2}$ is used here as a descriptive comparison
statistic rather than for calculating a formal goodness-of-fit
probability.

A point-to-arm statistic alone may preferentially favor a model
containing more numerous or more extended arm segments, because each
source has more opportunities to lie close to one of the predicted
curves. We therefore also quantified how much of the predicted arm
structure is supported by the observations. We adopted the spiral-arm loci and their azimuthal extents as defined in the original R19 and X23 models, retaining only arm segments within $R_{\rm GC}\leq15$ kpc. The resulting arm loci were then sampled at a common interval in arc length. For each sampled arm position
$\boldsymbol{c}_m$, we calculated the distance to the nearest observed
source,
\begin{equation}
    q_m =
    \min_i
    \left\|
        \boldsymbol{c}_m-\boldsymbol{x}_i
    \right\|.
\end{equation}

The supported-arm fraction was then defined as
\begin{equation}
    f_{\rm sup}
    =
    \frac{
        \sum_m \Delta l_m
        I\left(q_m\leq R_{\rm sup}\right)
    }{
        \sum_m \Delta l_m
    },
    \label{eq:arm_support}
\end{equation}
where $\Delta l_m$ is the arm length represented by the $m$th sampled
position and $R_{\rm sup}$ is the adopted support radius. A large value
of $f_{\rm sup}$ therefore indicates that a large fraction of the
predicted arm length has at least one nearby Cepheid. For the two models, taking into account the arm width ($\sim0.4$kpc) and the distance error of classical Cepheid variables ($\sim0.3$kpc), we adopt the same support radius (0.5kpc) and curve sampling interval.

In this definition, a sampled arm position is considered supported if any Cepheid in the complete parent sample lies within $R_{\rm sup}$, irrespective of its subsequent arm assignment or rejection. Consequently, in regions where neighboring arms are separated by less than or are comparable to the adopted support scale, a Cepheid may contribute to the supported lengths of more than one arm. The supported-arm fraction should thus be regarded as a descriptive measure of the spatial coverage of the predicted arm network by the currently available Cepheid sample.

The measured value of $f_{\rm sup}$ is also affected by the spatial completeness of the Cepheid sample. In particular, distant and highly extincted regions are less completely sampled and may therefore yield lower support fractions even when spiral-arm structures are present. Future infrared surveys are expected to provide additional Cepheids in these regions and enable a more complete evaluation of the distant arm segments.

For a convenient summary of the local and global statistics, we
additionally defined
\begin{equation}
    S =
    \widetilde{\delta}
    +
    \lambda\left(1-f_{\rm sup}\right),
    \label{eq:combined_score}
\end{equation}
where $\widetilde{\delta}$ is the median normalized distance and
$\lambda=1$ was adopted for the present comparison. A smaller value of
$S$ indicates a model with both small point-to-arm distances and a
large supported-arm fraction. This combined score is used only as a
descriptive ranking statistic, because its absolute value and the
relative contributions of its two terms depend on the adopted support
radius and weighting factor.

\section{Assigned Spiral Arm}
\label{secB}

Classical Cepheids in our sample have ages of several tens of Myr. Although they remain relatively young tracers of the Galactic disk, some Cepheids may possess sufficiently large peculiar motions to move away from their birth arms or to become spatially projected near another arm. Therefore, spatial coincidence with a modeled arm locus does not necessarily guarantee physical membership. 
The spiral arm structure of the Milky Way exhibits quasi-continuous and coherent features in the longitude$-$LSR velocity ($\ell-v_{\rm LSR}$), longitude$-$latitude ($\ell-b$), and Galactic plane ($X-Y$) distribution, reflecting its nature as a large-scale structure \citep[e.g.,][]{reid2019, xu2023}. Leveraging this property, we assign Cepheids to spiral arms by combining these three spatial and kinematic diagnostics. 

{\scriptsize
\begin{equation}
    P_{*}^{\rm arm} = \exp\left[-\left(\frac{(X_{*}-X_{\rm arm})^2 + (Y_{*}-Y_{\rm arm})^2}{\max(\sigma_{*}, W_{\rm arm})^2} + \frac{(V_{*} - V_{\rm arm})^2}{10^2}\right)\right],
\end{equation}}

where the subscript $*$ denotes the measured properties of the classical Cepheid, and $\rm arm$ denotes the corresponding model values of the spiral arm. Here, $W_{\rm arm}$ represents the arm width, and $\sigma_{*}$ is the distance uncertainty of the star. Radial velocities of classical Cepheids are taken from \emph{Gaia} DR3, and are converted from the heliocentric frame to LSR. 

The model $v_{\rm LSR}$ values were obtained differently for 
R19 and X23. For R19, we adopted the spiral-arm kinematic parameters 
provided by \citet{reid2019}. Because X23 primarily provides the spatial 
loci of the spiral arms and does not specify an independent 
$v_{\rm LSR}$ locus for every arm segment, we calculated the X23 
velocities from the Galactocentric radii and heliocentric distances of 
points along its arm loci. The calculation uses the Galactic rotation 
model and Galactic parameters adopted by \citet{reid2019}, thereby 
placing the X23 arm loci in the same kinematic framework as R19.

For the Cepheids used in our analysis, the median value of the Gaia DR3 \texttt{radial\_velocity\_error} is approximately $3~\mathrm{km\,s^{-1}}$. This formal uncertainty, which also reflects the scatter among the RVS measurements for the brighter sources, is smaller than the characteristic peculiar and streaming motions expected for young spiral-arm tracers. We therefore adopt $10~\mathrm{km\,s^{-1}}$ as the characteristic scale in the velocity term to account primarily for spiral-arm-related streaming motions, other non-circular motions, and uncertainties in the idealized arm velocity loci, rather than as the formal uncertainty of the Gaia radial velocities \citep{reid2019,ramonfox2018}.

For Cepheids with radial velocity measurements, the arm-association probability is calculated using both the positional and velocity offsets. If either offset exceeds three times its characteristic scale, we set $P_{*}^{\rm arm}=0$. For Cepheids without radial velocity measurements, the velocity term is omitted and the arm-association probability is calculated using only their positions in the $X$--$Y$ plane. In this case, $P_{*}^{\rm arm}=0$ if the positional offset exceeds three times its corresponding uncertainty. The arm with the highest probability is finally adopted as the arm associated with the Cepheid.

\section{Spiral arms fitting}
\label{secC}

We fit a spiral pattern to the spiral arm segments by adopting a log-periodic spiral, 
defined as,
\begin{equation}
    \ln (R / R_{\rm ref}) = -(\beta - \beta_{\rm ref}) \tan \psi,
    \label{eq:arms}
\end{equation}

where $R_{\rm ref}$ is the arm with
radius at reference azimuth $\beta_{\rm ref}$, and $\psi$ is pitch angle. 
We fit a straight line to $(x, y) = (\beta, \ln (R / R_{\rm ref}))$ using a Bayesian 
Markov Chain Monte Carlo (MCMC) procedure to estimate the parameters $R_{\rm ref}$ and $\psi$.
Considering that the finite ages of the stellar samples may introduce an intrinsic dispersion, $\delta_{\rm arm}$, in their positions relative to the spiral-arm centers, we incorporate this term into the positional uncertainties of the sources. Here, $\delta_{\rm arm}$ is treated as a free parameter in the fitting. 
The likelihood function is then defined as follows:
\begin{equation}
\ln \mathcal{L}
=
-0.5 \sum \left[
\frac{(R_{\rm obs} - R_{\rm model})^2}{R_{\rm sig}^2 + \delta_{\rm arm}^2}
+ \ln \left( R_{\rm sig}^2 + \delta_{\rm arm}^2 \right)
\right],
\end{equation}
where $R_{\rm sig}$ is the observation error.


\begin{thebibliography}{aasjournal7.1}

\bibitem[Anderson et al.(2016)]{anderson2016} Anderson, R.~I., M{\'e}rand, A., Kervella, P., et al.\ 2016, \mnras, 455, 4, 4231. doi:10.1093/mnras/stv2438

\bibitem[Anderson et al.(2024)]{anderson2024} Anderson, R.~I., Viviani, G., Shetye, S.~S., et al.\ 2024, \aap, 686, A177. doi:10.1051/0004-6361/202348400

\bibitem[Bian et al.(2024)]{bian2024} Bian, S.~B., Wu, Y.~W., Xu, Y., et al.\ 2024, \aj, 167, 6, 267. doi:10.3847/1538-3881/ad4030

\bibitem[Bland-Hawthorn \& Gerhard(2016)]{blandhawthorn2016} Bland-Hawthorn, J. \& Gerhard, O.\ 2016, \araa, 54, 529. doi:10.1146/annurev-astro-081915-023441

\bibitem[Bono et al.(2005)]{bono2005} Bono, G., Marconi, M., Cassisi, S., et al.\ 2005, \apj, 621, 2, 966. doi:10.1086/427744

\bibitem[Bovy et al.(2016)]{bovy2016} Bovy, J., Rix, H.-W., Green, G.~M., et al.\ 2016, \apj, 818, 2, 130. doi:10.3847/0004-637X/818/2/130

\bibitem[Brunthaler et al.(2011)]{brunthaler2011} Brunthaler, A., Reid, M.~J., Menten, K.~M., et al.\ 2011, Astronomische Nachrichten, 332, 5, 461. doi:10.1002/asna.201111560

\bibitem[Buta (2013)]{buta2013} Buta, R.~J.\ 2013, Secular Evolution of Galaxies, 155. doi:10.48550/arXiv.1304.3529

\bibitem[Buta et al.(2015)]{buta2015} Buta, R.~J., Sheth, K., Athanassoula, E., et al.\ 2015, \apjs, 217, 2, 32. doi:10.1088/0067-0049/217/2/32

\bibitem[Craig et al.(2025)]{craig2025} Craig, P., Chakrabarti, S., Pettitt, A.~R., et al.\ 2025, \apj, 988, 2, 217. doi:10.3847/1538-4357/ade232

\bibitem[Dobbs \& Baba(2014)]{dobbs2014} Dobbs, C. \& Baba, J.\ 2014, \pasa, 31, e035. doi:10.1017/pasa.2014.31

\bibitem[Drimmel et al.(2025)]{drimmel2025} Drimmel, R., Khanna, S., Poggio, E., et al.\ 2025, \aap, 698, A230. doi:10.1051/0004-6361/202451100

\bibitem[Elmegreen \& Elmegreen(1989)]{elmegreen1989} Elmegreen, B.~G. \& Elmegreen, D.~M.\ 1989, \apj, 342, 677. doi:10.1086/167628

\bibitem[Ge et al.(2024)]{ge2024} Ge, Q.~A., Li, J.~J., Hao, C.~J., et al.\ 2024, \aj, 168, 1, 25. doi:10.3847/1538-3881/ad5201

\bibitem[Ghez et al.(2008)]{ghez2008} Ghez, A.~M., Salim, S., Weinberg, N.~N., et al.\ 2008, \apj, 689, 2, 1044. doi:10.1086/592738

\bibitem[Grand et al.(2016)]{grand2016} Grand, R.~J.~J., Springel, V., Kawata, D., et al.\ 2016, \mnras, 460, 1, L94. doi:10.1093/mnrasl/slw086

\bibitem[Hao et al.(2021)]{hao2021} Hao, C.~J., Xu, Y., Hou, L.~G., et al.\ 2021, \aap, 652, A102. doi:10.1051/0004-6361/202140608

\bibitem[Hilmi et al.(2020)]{hilmi2020} Hilmi, T., Minchev, I., Buck, T., et al.\ 2020, \mnras, 497, 1, 933. doi:10.1093/mnras/staa1934

\bibitem[Hou \& Han(2015)]{hou2015} Hou, L.~G. \& Han, J.~L.\ 2015, \mnras, 454, 1, 626. doi:10.1093/mnras/stv1904

\bibitem[Hyland et al.(2026)]{hyland2026} Hyland, L.~J., Reid, M.~J., Ellingsen, S.~P., et al.\ 2026, \apj, 1004, 2, 209. doi:10.3847/1538-4357/ae64f5

\bibitem[Katz et al.(2023)]{katz2023} Katz, D., Sartoretti, P., Guerrier, A., et al.\ 2023, \aap, 674, A5. doi:10.1051/0004-6361/202244220

\bibitem[Kong et al.(2025)]{kong2025} Kong, D., Xu, Y., Liu, D., et al.\ 2025, \aj, 170, 4, 243. doi:10.3847/1538-3881/ae0320

\bibitem[Kormendy \& Kennicutt(2004)]{kormendy2004} Kormendy, J. \& Kennicutt, R.~C.\ 2004, \araa, 42, 1, 603. doi:10.1146/annurev.astro.42.053102.134024

\bibitem[Lemasle et al.(2022)]{lemasle2022} Lemasle, B., Lala, H.~N., Kovtyukh, V., et al.\ 2022, \aap, 668, A40. doi:10.1051/0004-6361/202243273

\bibitem[Minniti et al.(2021)]{minniti2021} Minniti, J.~H., Zoccali, M., Rojas-Arriagada, A., et al.\ 2021, \aap, 654, A138. doi:10.1051/0004-6361/202039512

\bibitem[Morgan et al.(1953)]{morgan1953} Morgan, W.~W., Whitford, A.~E., \& Code, A.~D.\ 1953, \apj, 118, 318. doi:10.1086/145754

\bibitem[Ram{\'o}n-Fox \& Bonnell(2018)]{ramonfox2018} Ram{\'o}n-Fox, F.~G. \& Bonnell, I.~A.\ 2018, \mnras, 474, 2, 2028. doi:10.1093/mnras/stx2866

\bibitem[Reid et al.(2009)]{reid2009} Reid, M.~J., Menten, K.~M., Zheng, X.~W., et al.\ 2009, \apj, 700, 1, 137. doi:10.1088/0004-637X/700/1/137

\bibitem[Reid et al.(2019)]{reid2019} Reid, M.~J., Menten, K.~M., Brunthaler, A., et al.\ 2019, \apj, 885, 2, 131. doi:10.3847/1538-4357/ab4a11

\bibitem[Skowron et al.(2025)]{skowron2025} Skowron, D.~M., Drimmel, R., Khanna, S., et al.\ 2025, \apjs, 278, 2, 57. doi:10.3847/1538-4365/adc3f3

\bibitem[Steiman-Cameron(2010)]{steiman2010} Steiman-Cameron, T.~Y.\ 2010, Galaxies and their Masks, 45. doi:10.1007/978-1-4419-7317-7\_3

\bibitem[VERA Collaboration et al.(2020)]{vera2020} VERA Collaboration, Hirota, T., Nagayama, T., et al.\ 2020, \pasj, 72, 4, 50. doi:10.1093/pasj/psaa018


\bibitem[Wang et al.(2018)]{wang2018} Wang, S., Chen, X., de Grijs, R., et al.\ 2018, \apj, 852, 2, 78. doi:10.3847/1538-4357/aa9d99

\bibitem[Wegg et al.(2015)]{wegg2015} Wegg, C., Gerhard, O., \& Portail, M.\ 2015, \mnras, 450, 4, 4050. doi:10.1093/mnras/stv745


\bibitem[Wei et al.(2024)]{wei2024} Wei, J., Xu, Y., Lin, Z., et al.\ 2024, \aj, 168, 6, 264. doi:10.3847/1538-3881/ad8632

\bibitem[Xu et al.(2006)]{xu2006} Xu, Y., Reid, M.~J., Zheng, X.~W., et al.\ 2006, Science, 311, 5757, 54. doi:10.1126/science.1120914

\bibitem[Xu et al.(2018)]{xu2018a} Xu, Y., Bian, S.~B., Reid, M.~J., et al.\ 2018a, \aap, 616, L15. doi:10.1051/0004-6361/201833407

\bibitem[Xu et al.(2018)]{xu2018b} Xu, Y., Bian, S.~B., Reid, M.~J., et al.\ 2018b, \aap, 616, L15. doi:10.1051/0004-6361/201833407

\bibitem[Xu et al.(2021)]{xu2021} Xu, Y., Hou, L.~G., Bian, S.~B., et al.\ 2021, \aap, 645, L8. doi:10.1051/0004-6361/202040103

\bibitem[Xu et al.(2023)]{xu2023} Xu, Y., Hao, C.~J., Liu, D.~J., et al.\ 2023, \apj, 947, 2, 54. doi:10.3847/1538-4357/acc45c

\bibitem[Xu \& Liu(2026)]{xu2026} Xu, Ye \& Liu, De-jian,\ 2026, \caa, 50, 2, 260. doi:10.1016/j.chinastron.2026.06.002

\end{thebibliography}
\end{document}